\documentclass[preprint]{aastex}
\usepackage{emulateapj5,graphicx,times,url}

\slugcomment{accepted by ApJ}

\shorttitle{Star Formation in the Tail}
\shortauthors{Yagi et al.}
\begin{document}

\title{Multi-Wavelength Studies of Spectacular 
Ram Pressure Stripping of a Galaxy. II.
Star Formation in the Tail}

\author{Masafumi Yagi\altaffilmark{1,2},
Liyi Gu\altaffilmark{3,4},
Yutaka Fujita\altaffilmark{5},
Kazuhiro Nakazawa\altaffilmark{3},
Takuya Akahori\altaffilmark{6},
Takashi Hattori\altaffilmark{7},
Michitoshi Yoshida\altaffilmark{8}, 
Kazuo Makishima\altaffilmark{3,4,9},
}

\altaffiltext{1}{email:YAGI.Masafumi@nao.ac.jp}
\altaffiltext{2}{
Optical and Infrared Astronomy Division,
National Astronomical Observatory of Japan,
2-21-1, Osawa, Mitaka, Tokyo, 181-8588, Japan}
\altaffiltext{3}{Department of Physics, The University of Tokyo, 
7-3-1 Hongo, Bunkyo-ku, Tokyo 113-0011, Japan}
\altaffiltext{4}{
Reserch Center for Early Universe, School of Science,
The University of Tokyo, 7-3-1 Hongo, Bunkyo-ku, Tokyo 113-0011, Japan}
\altaffiltext{5}{
Department of Earth and Space Science, Graduate School of Science, 
Osaka University, Toyonaka, Osaka 560-0043, Japan
}
\altaffiltext{6}
{
Sydney Institute for Astronomy, School of Physics, 
The University of Sydney, 
44 Rosehill Street, Redfern, NSW 2006, Australia
}
\altaffiltext{7}
{
Subaru Telescope, 650 North A'ohoku Place, Hilo, Hawaii 96720, USA
}
\altaffiltext{8}{
Hiroshima Astrophysical Science Center, Hiroshima University,
1-3-1, Kagamiyama, Higashi-Hiroshima, Hiroshima, 739-8526, Japan
}
\altaffiltext{9}
{
MAXI Team, Institute of Physical and Chemical Research, 2-1 Hirosawa,
Wako, Saitama 351-0198
}

\begin{abstract}

With multiband photometric data in public archives, we detected four
intracluster star-forming regions in the Virgo cluster.  Two of them
were at a projected distance of 35 kpc away from NGC~4388, and the
other two were 66 kpc away.  Our new spectroscopic observation
revealed that their recession velocities were comparable to the 
ram-pressure-stripped tail of NGC~4388 and confirmed their
association.  The stellar mass of the star-forming regions ranged
from $10^{4}$-$10^{4.5}$ $M_{\odot}$ except for that of the faintest one
which would be $<10^3$ M$_{\odot}$.  The metallicity was comparable to
the solar abundance, and the age of the stars was $\sim 10^{6.8}$ years.
Their young stellar age meant that the star formation should have
started after the gas was stripped from NGC~4388.  This implied 
in situ condensation of the stripped gas.  We also found that two
star-forming regions lay near the leading edge of a filamentary dark
cloud.  The extinction of the filament was smaller than 
that derived from
the Balmer
decrement of the star-forming regions.  It implied that the dust in
the filament would be locally dense around the star-forming regions.

\end{abstract}

\keywords{galaxies: individual(NGC4388) --- galaxies: intergalactic medium --- H II regions}

\def\Ha{H$\alpha $}
\def\Hb{H$\beta $}
\def\Hd{H$\delta $}

\def\HI{HI}
\def\HII{HII}

\section{Introduction}

In a cluster of galaxies, 
the galactic interstellar medium 
can be transferred into intergalactic space by various processes
\citep[and references therein]{Boselli2006}.  The fate of the gas is,
however, still not totally understood.  Some of the gas would be
cooled to form stars in intergalactic space; some would accrete
back to the parent galaxy or to other galaxies in the cluster and some
would be heated to become hot plasma in the cluster.

If the gas is heated to $\sim 10^7$ K, it would be observed in X-ray.
Actually, such X-ray tails associated with galaxies have been reported by
many studies
\citep[e.g.,][]{Iwasawa2003,Wang2004,Machacek2005,Sun2005,Sun2006,Fujita2006,Machacek2006,Sun2007,Sun2010,Wezgowiec2011,Gu2013}.
The gas in the ionization-recombination balance phase
would be observed in \Ha.  
Galaxies' \Ha\ ``tails'' have also been reported by many studies
\citep[e.g.,][]{Gavazzi2000,
Gavazzi2001,Yoshida2002,Sun2007,Yagi2007,Kenney2008,
Sun2010,Smith2010,Yagi2010,Fossati2012}.
If the gas remains cool, it would observed in radio waves as 
\HI\  tails \citep[e.g.,][]{Oosterloo2005,Chung2007,Vollmer2007}, or
molecular clouds \citep[e.g.,][]{Vollmer2008,Vollmer2012}.  
Several studies have presented models and 
and simulations of the evolution of the gas in galaxies
\citep[e.g.,][]{Roediger2006,Roediger2007,Roediger2008,Kepferer2008,Tonnesen2012}.

Note that the coexistence of the X-ray, \Ha\ and \HI\ 
phase in a tail is rare \citep{Roediger2009,Sun2010}.  NGC~4388 in the
Virgo cluster is one of the rare samples.  From deep \Ha\ imaging,
\citet{Yoshida2002} discovered a very extended ionized region near the
galaxy.  The region was spectroscopically observed 
with the Faint Object Camera and
Spectrograph \citep[FOCAS]{Kashikawa2002} at the Subaru Telescope
\citep{Yoshida2004}.  \citet{Yoshida2004} confirmed that the region
had a recession velocity comparable to that of NGC~4388.  They also argued
that the \Ha\ tail would have been made by ram-pressure stripping
\citep{Gunn1972,Fujita1999}.  \citet{Oosterloo2005} observed the
region in \HI\  21cm with the Westerbork Synthesis Radio Telescope
and found that the \HI\  tail extends far distant to $>120$ kpc
from NGC~4388.  
\citet{Kenney2008} discovered another group of \Ha\ clouds 
between NGC~4438 to the east and M86 to the west. 
The clouds have $\sim$ 0 
heliocentric recession velocity. 
Some part of the NGC~4438-M86 \Ha\ clouds overlap the 
extended gas of 
NGC~4388, but the difference of
the 
heliocentric recession velocity enables us to distinguish them.  And recently,
the association of an X-ray gas to the tail near NGC~4388 was reported
by \citet{Wezgowiec2011}.

The origin of the tail of NGC~4388 was discussed in several studies.
\citet{Vollmer2009} assumed that NGC~4388 is about 150~Myr
after the closest approach to the cluster center, i.e. M87, and 
calculated the tangential velocities and the line-of-sight distance.
They also mentioned the possibility that the tail was formed by
the ram pressure from the intracluster medium (ICM) of the M86 group.

In the first paper of this series \citep[hereafter paper I]{Gu2013},
we reported that the \HI\  gas accompanies the hot X-ray gas
even $\sim$ 100 kpc away from NGC~4388,
and NGC~4388 and its \HI\  tail lie in front of M86 system.
We compared the effect of the ICM of the M86 group and that of 
the Virgo cluster (M87), and concluded that the tail would have been 
made by ram-pressure stripping of ICM of the Virgo cluster.

In this paper, we report the identification of 
four intracluster/intergalactic star-forming regions 
in the Virgo cluster, which are associated with
the \HI\  tail of NGC~4388.
Intergalactic star formation in stripped gas from a galaxy 
was reported by several authors
\citep{Owen2006,Sun2007,
Cortese2007,Yoshida2008,Sun2010,Sivanandam2010,Smith2010,
Hester2010,Fumagalli2011,Abramson2011,
Fossati2012,Yoshida2012,ArrigoniBattaia2012,Boissier2012,Ohyama2013}.
Besides those in the tails, 
isolated star-forming regions in the Virgo cluster
were also reported
\citep{Gerhard2002,Arnaboldi2003,Cortese2003,Cortese2004}.
\citet{Kronberger2008} and 
\citet{Kapferer2009} simulated the ram-pressure stripping of a galaxy
and the star formation in the tail.
\citet{Kronberger2008} showed that 
a significant fraction of stars would be formed in the tail, 
and \citet{Kapferer2009} predicted that 
young stars would be detectable throughout the whole tail
up to 400 kpc.
\citet{Yamagami2011} showed that 
molecular clouds should be formed in the tail.
The study in this paper
would be directly compared with these model predictions.

The structure of this paper is as follows; In Section 2 and Appendix
A, we described the data we used, which includes a new spectroscopic
observation.  The spectroscopic data were analysed in Section 3, and
physical parameters of four \Ha\ emitting regions were derived.  The
nature of the regions was discussed in Section 4, and summarized in
Section 5.  We adopt the distance to NGC~4388 and its tail as 16.7 Mpc
\citep{Yoshida2002}. The distance modulus is m-M=31.11, and one arcsec
corresponds to 81.0 pc.  The magnitude system is AB-system
\citep{Oke1983} unless otherwise noted.

\section{Data}

\subsection{Imaging data}

For detailed analysis of the \Ha\ counterpart along the NGC~4388 tail,
we retrieved data obtained with the Subaru Prime Focus Camera
\citep[Suprime-Cam][]{Miyazaki2002} around the \HI\  tail 
in N-A-L659 filter 
\citep[NA659 hereafter]{Yoshida2002,Okamura2002,Hayashino2003}
and R (W-C-RC)-band filter.
The center wavelength of NA659 was 6600\AA\ and the full width at 
half maximum (FWHM) was 100\AA,
which corresponds to \Ha\ in recession velocity of 
$v$=1700$\pm$2300 km s$^{-1}$.
The NA659 imaging dataset was the same as that used in \citet{Yoshida2002},
but we reprocessed it from the raw data for this paper.
Recently, the region was observed much deeper in R-band.
We used the data for the continuum subtraction.
Also, V (W-J-V), i (W-S-I+), and N-A-L503
\citep[NA503 hereafter]{Yoshida2002,Okamura2002,Hayashino2003}
were available.
The center wavelength of NA503 was 5020\AA\ and FWHM was 100\AA.
NA503 covered [OIII]4959 and [OIII]5007 at 
recession velocity $v$=3680$\pm$3020, and $v$=780$\pm$3000 km s$^{-1}$,
and did not cover \Hb\ in the Virgo cluster.
The detail of the Suprime-Cam data reduction is given in Appendix A.
In the Appendix, the data in other wavelength are also described.

From eye inspection, we selected two fields 
for spectroscopic observation.
We call the four clumps of the \Ha\ emitting regions 
HaR-1, 2, 3, and 4 from the north to the south (Figure \ref{fig:VirgoFC}).
Their appearances 
are shown in Figures \ref{fig:stamps12} and \ref{fig:stamps34},
and coordinates are given in Table \ref{tab:pos}.
HaR-2 consists of several sub-clumps.
We call them HaR-2a, 2b, 2c and 2d, from the north to the south,
respectively. 
Their multiband magnitudes are given in Tables \ref{tab:photom} 
and \ref{tab:photom2}.
It should be noted that there are other \Ha\ emitting regions around 
the tail. The sampling of the \Ha\ emitting region in this study was 
not a complete selection.

\subsection{Spectroscopic data}

\subsubsection{FOCAS Observation and data reduction}

We carried out a longslit spectroscopy at 
the two fields (HaR-1,2 and HaR-3,4) on 2013-03-03(UT) with 
the Faint Object Camera and Spectrograph \citep[FOCAS]{Kashikawa2002}
attached to the Subaru Telescope.
We used the longslit of 0.8 arcsec width, and 300B grism without 
order-cutting filters. 
The dispersion was $\sim$ 1.38 \AA\ pixel$^{-1}$.
The pixel scale of the FOCAS data was 
0.207 arcsec along the slit.

The data reduction was done in a standard manner:
overscan was subtracted and flat-fielded.
Cosmic rays are removed by python implementation of 
LA Cosmic package \citep{vanDokkum2001}%
\footnote{\url{http://obswww.unige.ch/~tewes/cosmics_dot_py/}}.
The wavelength was calibrated using the sky emissions and 
the Thorium-Argon comparison lamp obtained during the observation. 
The measured instrumental profile was $\sim$ 9 \AA\ in FWHM.
Then, the sky emissions and the absorption features of sky background 
(mainly moonlights) are subtracted.
The 2D spectra around \Hb\ and \Ha\ are shown in Figure \ref{fig:2Dspec}.

Feige~67 was observed with longslit of 2.0 arcsec width
as a spectrophotometric standard. 
The calibrated spectrum was obtained from calspec
in STSci\footnote{\url{ftp://ftp.stsci.edu/cdbs/current_calspec/}}.
The atmospheric dispersion corrector (ADC) at Cassegrain focus was 
not available in the observation,
and wavelength-dependent positional shift of the target 
across the slit should occur.
In our observation condition,
the differential atmospheric dispersion was negligible for HaR-3,4, 
while $\sim$ 13\% loss of flux may exist in the worst case for HaR-1,2.
We therefore regard \Hb/\Ha\ of HaR-1 and 2
as possibly having suffered 13\% systematic error.

\subsubsection{Spatial distribution and apertures}

The spatial distribution of flux around \Ha, [NII]6584, and 
continuum is shown in Figure \ref{fig:xprof}.
\Ha\ and [NII] are not net emission line but 
include continuum flux.
In HaR-1 and 4, \Ha, [NII], and continuum profiles are similar. 
In HaR-3, the continuum shows a small gradient, 
which would be a contamination from the neighboring galaxy at 
the position x$\sim$ 21.

In HaR-2, the spatial profile was rather complicated. 
At x$\sim$24, \Ha, and [NII] has a peak, while the continuum does not.
At x$\sim$25, the continuum has a peak, and \Ha\ and [NII] shows a bump.
Around x$\sim$27, \Ha\ and [NII] shows a small peak, but the continuum was 
$\sim$0. 
Then a second peak of the continuum was seen at 
the position in the slit x$\sim$29, 
where \Ha\ and [NII] are not so strong as in the other part.
We therefore divide the aperture of HaR-2 into four
which correspond to the clumps in Figure \ref{fig:stamps12} bottom.
Then spectra were extracted for the \Ha\ peaks with a Gaussian weight
along the slit.
The spectra are shown in Figure \ref{fig:spec}.
In the Figure, the wavelength dependence of 
the instrumental and atmospheric throughput is corrected,
except for the possible slit loss.
The neighboring galaxy between HaR-3 and 4 was found to be
a background star-forming galaxy at z=0.118.

\subsubsection{Measurement of the line strengths}

The strength of \Ha, [NII]6584, [SII]6717,6731, [OIII]5007, 
and \Hb\ are measured by Gaussian fitting.
In $6560\AA<\lambda<6820\AA$, \Ha, [NII], and [SII] are fitted
by a multivariate least-square fitting.
We assume that the background continuum is constant in the fitting region, 
the redshift is the same for the lines,
and the FWHM of the lines is the same (dominated by 
instrumental profile).
The strength of [NII]6548 was fixed to be 1/3 of [NII]6584.
The estimated redshifts are shown in Table \ref{tab:redshift}.
In $4800\AA<\lambda<5100\AA$, \Hb\ and [OIII]5007 are measured
in the same way.  [OIII]4959 was assumed to be 1/3 of [OIII]5007.
At HaR-2c and HaR-2d, the lines are too weak and we could not fit well.

Since the seeing size was comparable to the slit width,
and the slit width was narrower than the spatial extent of the HaRs, 
the flux value from the spectra has little physical meaning,
but the flux ratios are meaningful.
The ratios are shown in Table \ref{tab:lineratio}.
The error of each parameter was estimated by 
a residual bootstrap method of 1000 realization,
except for \Ha/\Hb.
The error of \Ha/\Hb\ was estimated from the error of \Ha\ flux and 
that of \Hb\ flux by the residual bootstrapping, assuming that
their errors were independent.
For HaR-1 and HaR-2, 13\% error due to possible differential 
atmospheric dispersion effect was also included.

In Figure \ref{fig:lineratios}, diagnostic line ratio plots 
are shown.
All the regions have a line ratio similar to an \HII\ region.
Moreover, [OI]6300 was not detected in any of the regions, 
possible contribution of shock as seen in
\Ha\ knots in \citet{Yoshida2004} are negligible in our targets.
Note that [OI]6300 of the HaRs are not affected by
the atmospheric [OI]6300 or [OI]6363, because of the redshift.

\section{Results}

\subsection{Redshift and position}

\begin{figure}
\includegraphics[scale=0.45,bb=0 0 1060 796]{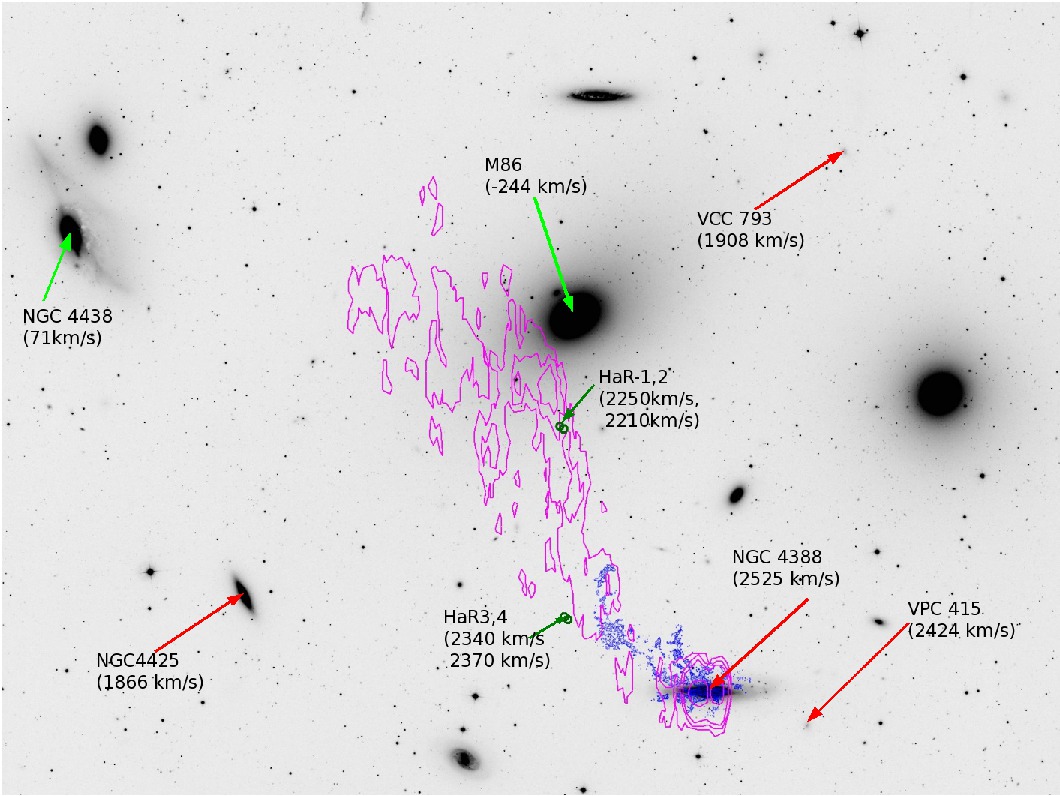}
\caption{Virgo cluster image in SDSS-r band,
coadded using NASA/IPAC IRSA Montage service.
North is up and east is to the left, and the size is 48$\times$36 arcmin.
Positions of targets in this study (HaRs) and marked as 
open dark green circles and indicated by arrow with labels.
The velocities in the labels are the heliocentric recession velocity.
Blue contours near NGC~4388 show the 
\Ha\ tail extracted from \Ha-R data.
Purple contours show the \HI\  distribution by \citet{Oosterloo2005}.
Four high-velocity galaxies are indicated by red arrows.
M86 and NGC~4438 are indicated by light green arrows.
}
\label{fig:VirgoFC}
\end{figure}

The HaRs have large ($>$2000 km s$^{-1}$) recession velocity.
The recession velocity of NGC~4388 is also large; 2524 km s$^{-1}$
\citep{Lu1993}, and that of the \HI\  tail is 2000 -- 2550 km s$^{-1}$
\citep{Oosterloo2005}.
As the recession velocity of the Virgo cluster is 
1079 km s$^{-1}$ \citep{Ebeling1998},
the peculiar velocity of NGC~4388 and its tail is $>900$ km s$^{-1}$.
In the 48$\times$36 arcmin region shown in Figure \ref{fig:VirgoFC},
the number of galaxies 
which have recession velocity larger than 1500 km s$^{-1}$
in NASA/IPAC Extragalactic Database(NED) is seven.
Two of them, AGC~226080 and VCC~956, have HI redshift only.
As the beam of the HI observation is large enough 
\citep[3.3$\times$3.8 arcmin;][]{Giovanelli2007}
to include 
the HI tail of NGC~4388, the high velocity would be a measurement of 
the tail by an accidental overlap.
The recession velocity of VCC896 is uncertain,
since it is inconsistent among
\citet{Conselice2001}, 
Sloan Digital Sky Survey Data Release 7
\citep[SDSS2 DR7;][]{DR7}%
\footnote{\url{http://cas.sdss.org/dr7/en/}}
and SDSS3 DR9
\citep{DR9}%
\footnote{\url{http://skyserver.sdss3.org/dr9/}}.
The rest four have optical spectroscopic redshift
and consistent among literatures.
We marked the four in Figure \ref{fig:VirgoFC}.
The HaRs are 
very likely to be associated with the \HI\ tail of NGC~4388,
because of the large resession velocity and their position;
HaR-1 and HaR-2 overlap the \HI,
and HaR-3 and HaR-4 lie near the contour edge (Figure \ref{fig:VirgoFC}),
and their nearest galaxy with large recession velocity is NGC~4388.

\subsection{Size and morphology}

HaR-1 is a point source, and was only recognized in narrowband.
It is marginally detected in NUV, and in optical broadbands 
which include strong emissions.
Since the seeing size in the NA659 was about 0.75 arcsec, which corresponds
to 60 pc, the size of HaR-1 would be smaller than 60 pc.

HaR-2 shows a blobby and elongated appearance.
The size was 530$\times$ 210 pc.
HaR-2ab and HaR-2d were recognized separately in optical and infrared 
images, while they were merged in UV image 
due to the 6 arcsec resolution of 
Galaxy Evolution Explorer (GALEX),

HaR-3 had a slightly elongated shape toward HaR-4 and showed a
sign of a tail with a blob. 
The tail was contaminated 
by the light from the neighboring background galaxy,
and the length of the tail was uncertain.
The size of HaR-3 including the tail was $>$300 $\times$ 180 pc.

HaR-4 showed a possible multi-core or elongated core
surrounded by a blob and extended plume 
(Figure \ref{fig:stamps34}).
The size including the plume was 380 $\times$ 360 pc.
It is interesting that HaR-3 had an \Ha\ tail toward 
NGC~4388.
In HaR-2ab and HaR-4, tail/plume was on the far side of 
NGC~4388.
The distance between HaR-3 and HaR-4 is 13 arcsec (1 kpc),
and the difference of the recession velocity of HaR-3 and 4 is 
30km s$^{-1}$. 
Even if the mass of HaR-3 and 4 are $\sim 10^6$ M$_{\odot}$,
they would not be bound gravitationally.

\subsection{Internal extinction and \Ha\ luminosity}

Internal extinction was estimated from the Balmer decrement 
\Hb/\Ha\ in our FOCAS spectra.
We adopted the formula by \citet{Calzetti1994},
\begin{equation}
E(B-V)\simeq 0.935 \: \ln((H\alpha/H\beta)/2.88).
\end{equation}
The estimated internal extinction is shown in Table \ref{tab:Haflux}.

In our observation, the seeing size was comparable to the slit width.
We therefore did not try the absolute spectrophotometric calibration
from the spectra only.
Instead, the spectra were used to convert 
NA659 magnitude to the total flux of \Ha.
The extinction corrected \Ha\ luminosity are given 
in Table \ref{tab:Haflux}.

\subsection{Metallicity}

\citet{Kewley2002} presented model curves to estimate 
oxygen abundance (log(O/H)+12) 
using several combinations of emission lines
with various values of the ionization parameter ($q$).
We can use [NII]/[SII], [NII]/\Ha, and [NII]/[OIII] to
estimate the metallicity.
[NII]/[OIII] was corrected for the internal extinction using \Hb/\Ha,
and other two ratios were not corrected.
$q$ was also estimated from the curves.

From the fitting, the metallicity of the regions was
estimated as log(O/H)+12$\sim$8.6--8.7.
It was comparable to the solar abundance \citep{Grevesse2010}.
\citet{Yoshida2004} reported that 
the metallicity of \Ha-emitting filaments near NGC~4388
was almost at the solar value.
Their ionization parameters correspond to
log(U)=-2.6, -3.5, -3.2 and -3.5, for HaR-1, 2a, 3, and 4, respectively,
where $U$ is the ratio of the ionization photon density
to the electron density. 
Since log(U) decreases with age of the \HII\ region,
relatively high value 
of HaR-1 implies that HaR-1 would be younger.

\subsection{Electron density}

From the Figure 5.8 of \citet{Osterbrock2006},
\citet{ODell2013} 
derived the formula to estimate the electron density ($n_e$)
\begin{equation}
\log (n_e) = 4.705 -1.9875 \: \left( {\rm [SII]6716/[SII]6731} \right),
\end{equation}
which was applicable in 0.65$<$[SII]6716/[SII]6731$<$1.3.
In this study, the error of [SII]6717/[SII]6731 was large, 
and all except HaR-1 was consistent with $n_e \lesssim 10^2$ cm$^{-3}$.
HaR-1 may have a higher electron density $n_e \sim 10^3$ cm$^{-3}$.

\subsection{Mass and Age estimation}
\label{sec:mass-age}

We calculated model magnitudes and \Ha\ flux at different age 
and compared them with the observation.
The detail of the model is given Appendix \ref{sec:agemassdetail}.
We plotted the model age versus the model mass 
to reproduce the observation in Figure \ref{fig:SB99}. 
The abscissa is the logarithmic age from the burst.
The ordinate is the required stellar mass of the HaR 
to reproduce the observed magnitude/luminosity at each age.
Because the time evolution of the \Ha\ luminosity, UV, optical 
and MIR magnitudes are different, the crossing point of 
the tracks will indicate the stellar mass and the age.

\Ha, IR, and UV were comparable around log(T)=6.8 and log(mass)=4.3 
in Geneva model for HaR-2, HaR-3, and HaR-4,
while i has an offset by $\sim$ 0.4 dex, which corresponds to
$\sim$1 magnitude and a factor of  $\sim$2.5.
Though the consistency in the Padova model
was not as good as that of the Geneva model,
the best-fit mass and age were comparable.

The disagreement of the estimated mass and age among the data
might be explained partly by the difference in aperture sizes.
As UV resolution was worse, it required the largest aperture (r=6 arcsec),
and may include contaminants, which made the observed luminosity larger
and shifted the lines in Figure \ref{fig:SB99} upward.
However, IR had a smaller aperture (r=3 arcsec) than UV,
and \Ha\ is based on NA659 photometry, which was aperture 
photometry of 2.5 $\times$ Kron radius ($rK$). 
The 2.5$rK$ was 2$\pm$0.3 arcsec in the data, 
and therefore it was the smallest aperture, because of the best seeing.
Suprime-Cam i-band and 
Kitt Peak National Observatory (KPNO) mosaic ha-band 
were also measured in an aperture 
of 2.5$rK$ radius, which was 2--3 arcsec.
Nevertheless, \Ha, UV, and IR showed a good agreement, while
i-band and KPNO ha-band showed an offset in HaR-2 and HaR-3.
As the i-band magnitude of Suprime-Cam was comparable 
to SDSS photometry for HaR-4, which was based on Petrosian radius,
and the KPNO photometry of ha-band showed a similar trend, 
the disagreement would not be caused by photometric error.

The disagreement also could come
from model uncertainties and adopted assumptions, 
such as initial mass function (IMF), 
instantaneous star formation, and/or dust extinction model.
A possible model is to divide the region into
several sub-regions and to assume different extinctions.
For example, if some part has a large extinction,
UV is suppressed and optical and IR fluxes could be 
larger relative to UV and \Ha.
In fact, HaR-2 consisted of sub-clumps with various extinctions.
Though we can consider various star-formation histories and models,
it is beyond the scope of this study.
We do not go further to tune the star-formation histories 
and the models.

In Figure \ref{fig:SB99}, HaR-1 shows large disagreement 
of the mass and the age among the data.
Since HaR-1 would have a smaller mass, 
which was expected from fainter i-band magnitude,
it may have suffered a stochastic effect on IMF \citep[e.g.,][]{Koda2012}.

As shown in Table \ref{tab:redshift},
HaR-1 and 2 are 66 kpc and HaR-3 and 4 are 35 kpc away from NGC~4388.
\citet{Oosterloo2005} and \citet{Vollmer2009} 
assumed that the age of 
the \HI\ cloud is $\sim$ 200~Myr, and estimated the 
projected velocity of NGC~4388 to the cloud was 500 km s$^{-1}$
If we adopt the value as the relative velocity of HaRs to 
NGC~4388, it should take 130~Myr (HaR-1 and 2) and 
70~Myr (HaR-3 and 4) to reach the distance.
Since the time-scales are longer than 
the stellar age estimated above,
it suggests that the star formation in HaRs 
began after the gas left the host galaxy.
This is consistent with the prediction by \citet{Yamagami2011}.

\section{Discussion}

\subsection{Comparison with previous studies}

Table \ref{tab:SFregions} summarizes the literature concerning
a stripped tail in nearby clusters of galaxies.
Since the environment and physical condition of the tails
(the ambient gas temperature and density, the relative velocity,
 the density of the tail, etc) shows a variety. 
we cannot compare them directly. We can at least say that
HaR-1 and 2 are the most distant star-forming regions in the table,
and HaR-1 may be the least massive region.
Also, we can say that $10^{4-5}$ M$_{\odot}$ and $\sim 10^7$ yr 
star-forming regions in a stripped tail are not peculiar objects.

The age of the regions in \citet{Fumagalli2011} is 
substantially larger than the others
probably because a different model of the 
star-formation history was adopted.
The mass estimation is not affected by the different star-formation history.
The detection of the star-forming regions was based on \Ha\ 
and/or UV emissions, and therefore the sampling 
should be biased to the younger age.

\subsection{Comparison with \HI\ distribution and recession velocity}

The detailed heliocentric velocity distributions of \HI\ 
\footnote{Provided by Dr. J. van Gorkom.
The beam of the data was elongated from north to south:
77 arcsec in declination and 24.5 arcsec in right ascension.}
showed that the intensity of the \HI\ at the HaR-1,2 field 
had a wide peak around the recession velocity 
$\sim$ 2240 km s$^{-1}$ (intensity-weighted mean)
with width $\sim$ 150 km s$^{-1}$.
HaR-1 and 2 are just on the peak.
This means that the star-forming regions had only small 
peculiar velocity relative to the ambient \HI\  cloud along the line of sight.
Meanwhile, HI at HaR-3,4 field showed no significant peak. 
Since the star formation in stripped tail is thought to occur 
in the highest-density regions \citep{Kapferer2009},
it is puzzling that the \HI\  flux at HaR-3,4 field was low.

As we reported in paper I, the tail of NGC~4388 is 
accompanied by a relatively cool X-ray gas,
and the distribution of X-ray emission showed 
a marginal enhancement at the downstream of HaR-3,4.
The \HI\  gas around HaR-3 and 4 might already have evaporated,
except for the densest region. 
Another possibility is that such dense clouds are less affected by
ram pressure, and left behind the \HI\  gas \citep{Yamagami2011}.

\subsection{Positional offset of \Ha\ and stars, and dark cloud around HaR-1 and HaR-2}

The 2D spectrum and the slit profile of HaR-2 
(Figures \ref{fig:2Dspec},\ref{fig:xprof})
shows that the \Ha-strong region (HaR-2a)
appears left to the continuum regions (HaR-2b).
They resemble the {\it fireballs} found 
in the ram-pressure stripped tail in RB199 
in the Coma cluster \citep{Yoshida2008,Yoshida2012}.

If the gas cloud (HaR-2a, 2c) is now free from the ram pressure,
the gas and stars would move similarly.
On the other hand, because the ram pressure affects differently 
the stars and the gas,
it would cause a differential movement 
between the gas and the stars.
The positional shift between the gas and the stars implies 
that ram-pressure deceleration is still working here, 
66 kpc away from the host galaxy. 
Another possible explanation for the offset 
is that the gas has been consumed 
from the leading edge of the cloud, and the star formation
is propagated to the following regions.

In HaR-4, a plume also existed on the other side of the parent galaxy.
In HaR-3, however, the tail appeared on the side of the parent galaxy.
The direction was coincident to HaR-4.
The projected distance between HaR-3 and HaR-4 was about 1.0 kpc
and the difference of the recession velocity was 30 km s$^{-1}$.
If their mass is $\sim 10^5$ M$_{\sun}$, the velocity difference 
is so large that they cannot be gravitationally bound.
If HaR-3 and HaR-4 
had experienced an encounter in the past,
it means that such a small-scale motion existed in the tail.

Illuminated by the stars of the M86 envelope, 
a dark filamentary cloud was recognized around HaR-1 and HaR-2.
A high contrast image in V-band is shown as Figure \ref{fig:DC}.
In optical bands of Figure \ref{fig:stamps12}, 
the cloud was also recognized.
The dark cloud shows an elongated and twisting shape.
The length was $\sim$110 arcsec (7 kpc) and the width was 
$\sim$7 arcsec (600 pc) around HaR-2.

It is uncertain whether 
the dark cloud is associated with the NGC~4388 tail, 
an accidental overlap of a dark cloud of M86, 
or even if it is another dark cloud.
However, as such filaments of the dark cloud are only recognized 
along the \HI\  tail of NGC~4388, it is likely that 
the dark cloud is associated  with the NGC~4388 \HI\  tail.
These filaments of the dark cloud do not show any \Ha\ emitting 
regions except that with HaR-1 and HaR-2.
As there are only few star-forming regions along the tail,
it is unlikely that the dusts were created in the tail,
and the origin of the dusts should be NGC~4388.
It implies that the dusts were affected by the ram-pressure.
The ram-pressure stripping of dusts from disk of galaxies 
in the Virgo cluster was suggested by several authors;
NGC~4402 by \citet{Crowl2005},
NGC~4438 by \citet{Cortese2010b},
and NGC~4330 by \citet{Abramson2011}.
\citet{Cortese2010a} studied HI-deficient galaxies in the Virgo cluster,
and found that dust is stripped in the cluster environment 
as well as gas.
\citet{Yoshida2012} demonstrated 
the coincidence of the ram-pressure stripping of HI, 
\Ha\ and the dust in IC~4040.
The elongated and twisting shape of the dark cloud 
downstream of HaR-1 and HaR-2 may also 
be explained as a result of the ram pressure
which is still effective around the regions.

In paper I, we showed evidence that 
M86 is more distant than the NGC~4388 tail.
The total extinction of the cloud was estimated by
assuming that the background M86 light was smooth.
The estimated extinction at a region between HaR-1 and HaR-2
was $\sim$ 15\% and $\sim$ 10\% in V-band and in R-band,
respectively. 
It corresponds to $E(B-V)\sim 0.05$.
As the $E(B-V)$ estimated from the Balmer decrement in HaR-1 is 
much larger as $E(B-V)$= 0.59$^{+0.13}_{-0.15}$,
it is suggested that 
the dust in the dark filament would be 
locally dense around the star-forming regions.

\subsection{Star formation in the clouds}

Based on the argument by \citet{Yamagami2011}, we discuss the 
star formation in the molecular clouds that later shine as 
the observed \Ha\ emitting regions.
\citet{Elmegreen1997}
derived the relation between the mass of a cloud $M_c$ and the 
star-formation efficiency $\epsilon$ when 
the cloud is under pressure $P_c$.
We estimate in paper I that the ram-pressure on
NGC~4388 and the clouds should be $P_{\rm ram}\sim 
4\times 10^{-12}\rm\: dyn\: cm^{-12}$. In section \ref{sec:mass-age}, 
we found that the total mass of stars in
each \Ha\ emitting region is 
$M_{*}\sim 10^{4}$--$10^{4.5}\: M_{\odot}$. Assuming that
$P_c=P_{\rm ram}$, the mass of the molecular cloud before the stars were
born is $M_c \sim 10^{5.5}\: M_{\odot}$ because $\epsilon\sim 0.1$ (lower
part of Figure~4 in \citealt*{Elmegreen1997}). For this mass and pressure, 
the time-scale of star formation in the cloud is $\tau_{\rm form}\sim
10^7$~yr (upper part of Figure~4 in \citealt*{Elmegreen1997}; they referred to
the time-scale as 'disruption time'), which is consistent with the age of
the stars in the regions ($\sim 10^{6.8}$~yr).

\citet{Yamagami2011} also discussed the disruption of the clouds by
Kelvin-Helmholtz (KH) instability via interaction with the ambient
ICM. The relation among $M_c$, $P_c$ and the cloud radius $R_c$ before
star formation starts is given by
\begin{equation}
 \frac{M_c}{R_c^2}\sim 190\:{\rm M_{\odot}\: pc^2}
 \left(\frac{P_c}{1.38\times 10^{12}\rm\: erg\: s^{-1}}\right)^{1/2}
\end{equation}
\citep{Elmegreen1989}. For $M_c$ and $P_c$ we estimated, the cloud radius was
$R_c\sim 43$~pc. The time-scale of the development of KH instability is
\begin{equation}
 \tau_{\rm KH}=\frac{(\rho_c+\rho_{\rm ICM})R_c}
{(\rho_c\rho_{\rm ICM})^{1/2}v_c}
\end{equation}
where $\rho_c=3M_c/(4\pi R_c^3)$ is the density inside the cloud, and
$v_c$ is the relative velocity between the cloud and the ICM
\citep{Murray1993}. 
We estimate in paper I that $\rho_{\rm
ICM}\sim 2.1\times 10^{-28}\rm\: g\: cm^{-3}$, and thus we obtain
$\tau_{\rm KH}\sim 1.5\times 10^{7}$~yr. Since $\tau_{\rm KH} \sim
\tau_{\rm form}$, the cloud could marginally produce the stars before KH
instability develops. This may indicate that no magnetic fields are
required to protect the cloud from the development of KH instability
\citep{Yamagami2011}.

\subsection{Nature of HaR-1}

From the \Ha\ luminosity, 
the \Ha\ ionizing photon emission rate of HaR-1 was 
10$^{47.41}$ s$^{-1}$.
Assuming case-B recombination,
the logarithm of the total hydrogen ionizing photon emission rate $Q_H$ was
$log(Q_H) \sim 47.8^{+0.15}_{-0.13}$.
According to \citet{Sternberg2003},
a single B0V dwarf is enough to produce the photon rate,
and other stars such as B giants and O dwarfs would make too much \Ha.

However, a single B0V star is insufficient to 
reproduce the observed NUV magnitude.
The absolute V magnitude of B0V star is -4
\citep[e.g.;][]{Wegner2006}.
The NUV-V color of B0V was calculated with 
empirical spectral energy distributions (SEDs) 
by \citet{Pickles1998} as -1.0 mag.
The absolute NUV magnitude is therefore $\sim -5$.
Meanwhile, the absolute magnitudes of HaR-1 
converted from observed NUV magnitudes 
and the internal extinction correction was
-8.8$\pm$0.7 mag.
The observed UV magnitudes were $>3$ mag brighter.

The UV-optical color was also inconsistent with the following models.
We calculated the $NUV-i$ color of various SEDs,
and found that the bluest stars are O9V--B0V (NUV-i$\sim$-1.7).
The observed lower limit of 
NUV-i color of HaR-1 was $NUV-i<-1.6$, 
and if we correct the internal extinction of
$E_{\rm gas}(B-V)=0.59$, the color is $NUV-i<-2.9$.
The model SEDs, internal extinction estimated from the Balmer decrement,
the extinction law, and the factor of 0.44 were 
therefore inconsistent with the observed color.

\section{Summary}

In the \HI\  tail of NGC~4388 in the Virgo cluster, 
we identified four star-forming systems at a projected distance of 
35 kpc and 66 kpc away from the parent galaxy.

Main results are as follows:

\begin{enumerate}
\item The line ratios show typical \HII\ regions of solar metallicity,
with recession velocity comparable to that of \HI\ tail, 
which supports the idea that they are associated with the tail.
\item The \Ha\ luminosity and multiband photometry are 
fit with a burst model (STARBURST99), and we obtained reasonable solutions
within 0.3 dex. From the fitting,
the mass is 10$^{4-4.5}$ M$_{\odot}$ and the age is $\sim 10^{6.8}$ years,
except for the faintest one (HaR-1). The faintest one would have mass smaller 
than $10^3$ M$_{\odot}$, and was not fitted well, possibly because of 
the stochastic effect.
\item 
Their young age and large projected distance means that 
they were formed after they were removed from the parent galaxy.
This supports the theoretical prediction by \citet{Yamagami2011}.
The estimated age and mass, and the ram-pressure estimated in 
paper I
suggested that a magnetic field to protect the cloud 
from Kelvin-Helmholz instability \citep{Yamagami2011}
may not be necessary for the clouds.
\item One of the regions at 66 kpc away from NGC~4388 (HaR-2) 
shows an offset of \Ha\ emission from stars (fireball feature). 
This implies that ram-pressure is still effective here.
\item  Two of the regions (HaR-3,4) exist 
out of \HI\ distribution.
Evaporation of 
the \HI\  and/or the different deceleration by ram-pressure between 
the \HI\  gas and the condensed region may explain the result.
\item Two of the regions (HaR-1,2) lay at a leading edge of 
a filamentary dark cloud.
The cloud would have been stripped from NGC~4388 by the ram-pressure.
The extinction of the dark cloud is smaller than the internal extinction 
of the regions. The dust in the dark cloud would be 
locally dense around the star-forming regions.
\end{enumerate}

As the spectroscopic observation of this study was not a complete
sampling of the \Ha\ emitting regions along the tail, we did not 
investigate the global features in the tail 
such as total star-formation rate.
A future complete survey of the candidates of 
the star-forming region will
quantitatively set the constraints on the star-formation models 
in the tail.

\acknowledgments

We thank Drs. Tom Oosterloo and Jacqueline van Gorkom for
kindly providing us their map of \HI\  flux and 
the detailed \HI\  recession velocity distribution around our targets.
We acknowlege the referee Dr. Giuseppe Gavazzi for his 
suggestions and comments.
This work is based on observations obtained with the Subaru Telescope.
We acknowledge Subaru staff for support of the observation.
This work has made use of the 
SDSS-III database, NED database,
SMOKA archive,
STARS/MASTARS archive\footnote{\url{http://stars2.naoj.hawaii.edu/}},
NOAO science archive,
Mikulski Archive for Space Telescopes (MAST),
NASA/IPAC Infrared Service Archive (IRSA),
MPA-JHU DR7 release of spectrum measurements%
\footnote{\url{http://www.mpa-garching.mpg.de/SDSS/DR7/}}, 
Montage service by IRSA\footnote{
funded by the NASA's Earth Science Technology Office, Computational Technnologies Project, under Cooperative Agreement Number NCC5-626 between NASA and the California Institute of Technology. The code is maintained by the NASA/IPAC IRSA.},
and computer systems at Astronomy Data Center of NAOJ.
YF is supported by KAKENHI (23540308).

\begin{figure}
\includegraphics[scale=0.4,bb=0 0 998 778]{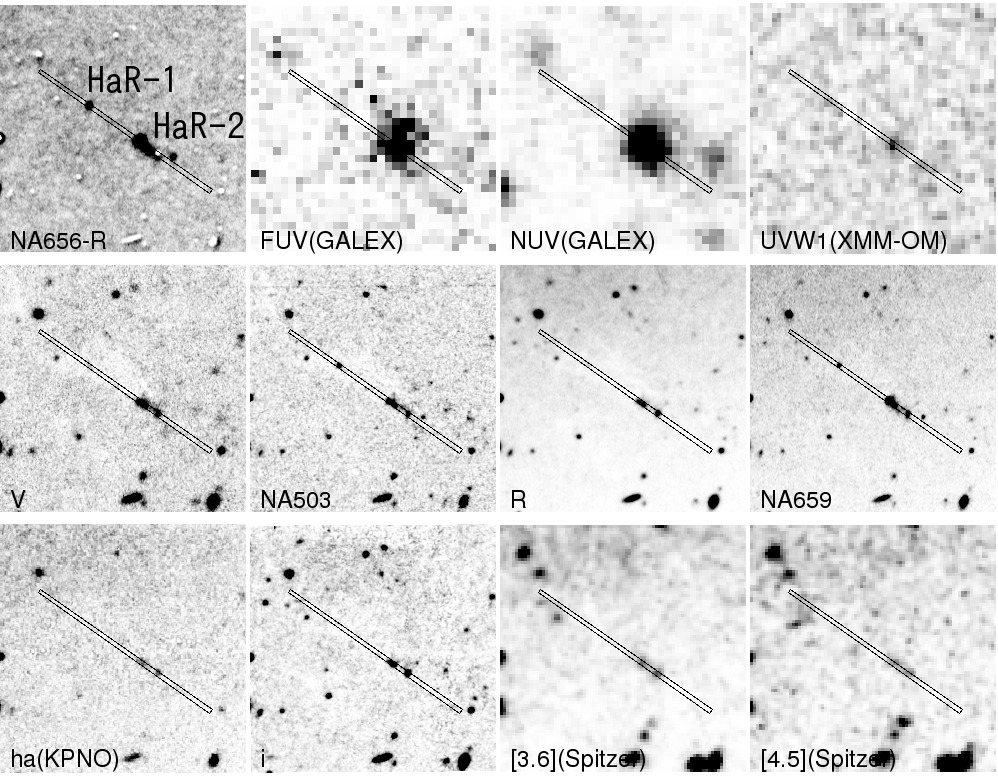}\\
\includegraphics[scale=0.6,bb=0 0 397 397]{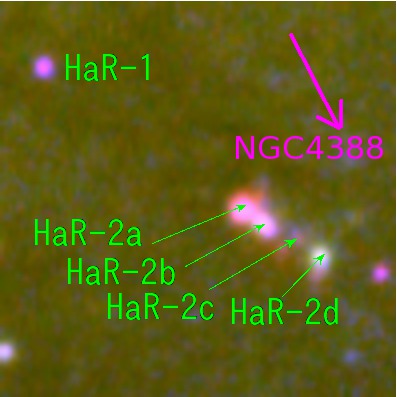}
\caption{(top)
Postage stamps of HaR-1 and HaR-2 in various wavelengths.
North is up, east is left, and the size is 50 arcsec square.
The filter and telescope name are given in label.
Those without the telescope name are the data of Suprime-Cam/Subaru.
The gray scale is arbitrary.
Slit position of FOCAS observation is overlaid.
(Bottom)
Three color (NA659, R, NA503) 
composite images. North is up, east is left, 
and the size is 20 arcsec square.
Color scale is arbitrary.
Magenta (red+blue) shows strong \Ha\ and [OIII] emissions 
at Virgo recession velocity.
The direction to the parent galaxy (NGC~4388) is shown by 
a magenta arrow at top-right.
}
\label{fig:stamps12}
\end{figure}

\begin{figure}
\includegraphics[scale=0.4,bb=0 0 998 778]{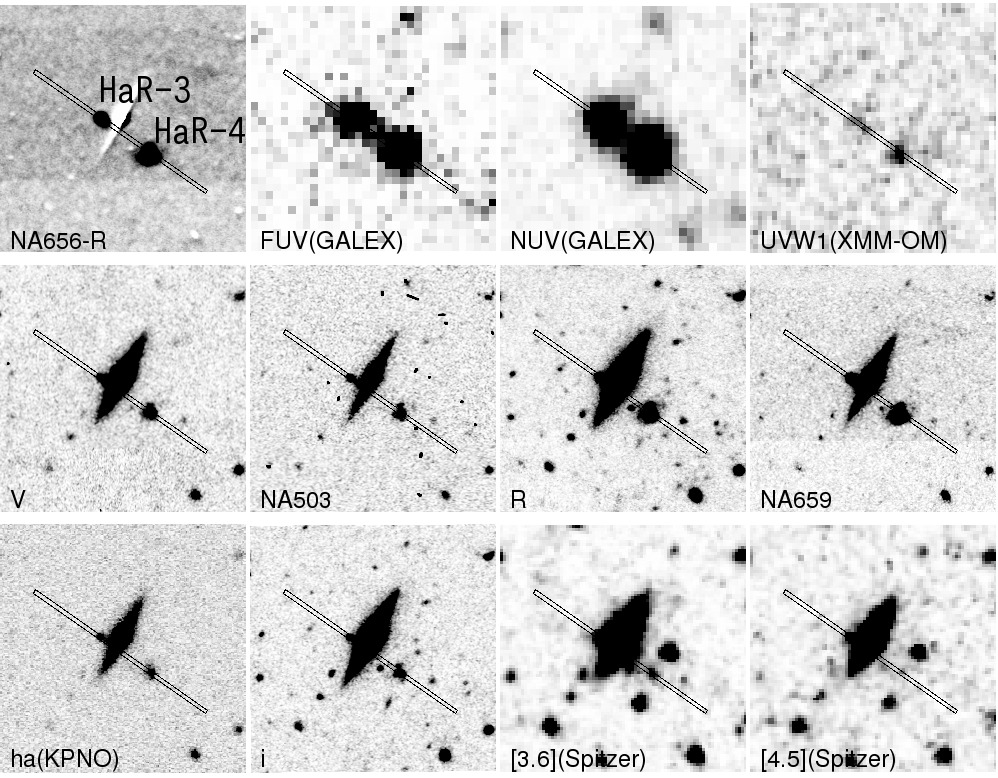}\\
\includegraphics[scale=0.6,bb=0 0 397 397]{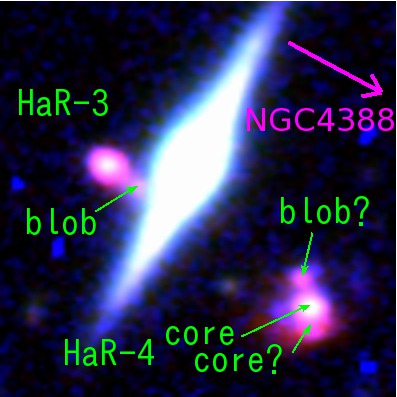}
\caption{
Same as Figure \ref{fig:stamps12}, but for HaR-3 and 4.
The apparent difference of background color of bottom panel 
of Figure \ref{fig:stamps12} and \ref{fig:stamps34}
is due to the M86 stars behind  HaR-1,2 field.
}
\label{fig:stamps34}
\end{figure}

\begin{figure}
\includegraphics[scale=0.5,bb=0 0 868 938]{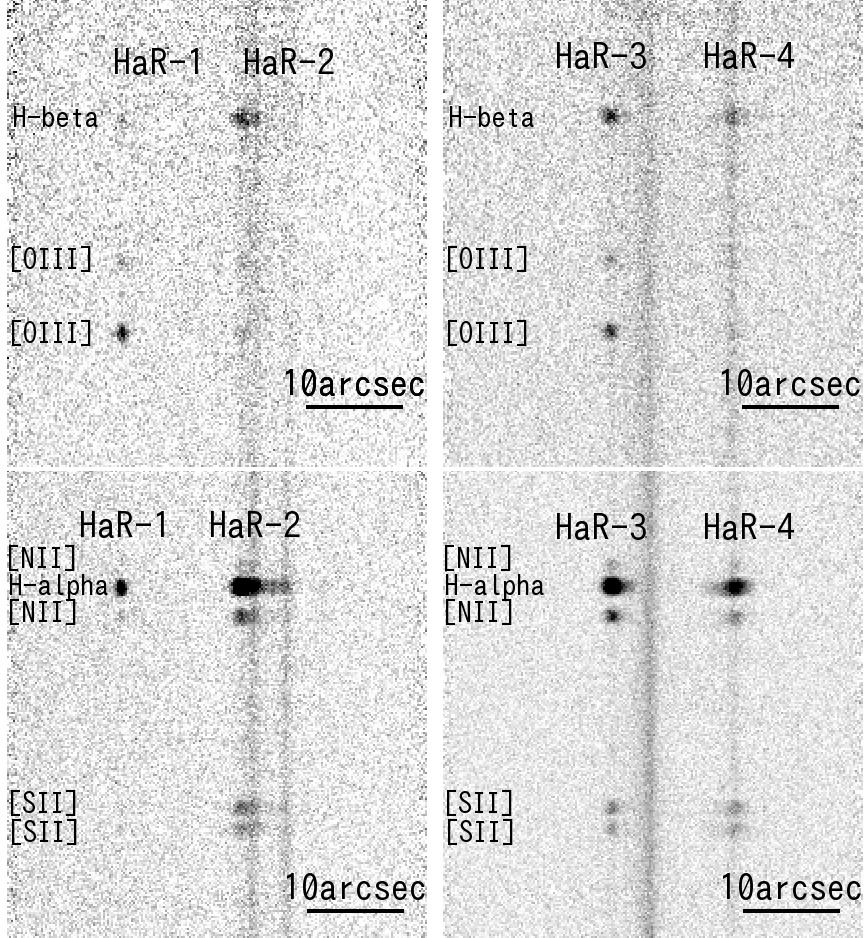}
\caption{Cutouts of FOCAS 2D spectra after sky subtraction.
Top panels show the spectra around \Hb\ and [OIII].
Bottom panels show the spectra around \Ha, [NII], and [SII].
}
\label{fig:2Dspec}
\end{figure}

\begin{figure}
\includegraphics[scale=0.3,bb=0 0 768 574]{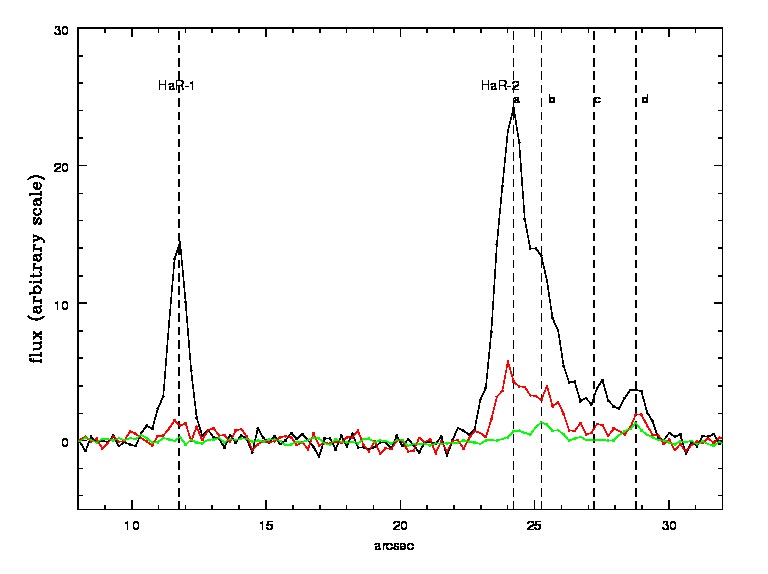}~
\includegraphics[scale=0.3,bb=0 0 768 574]{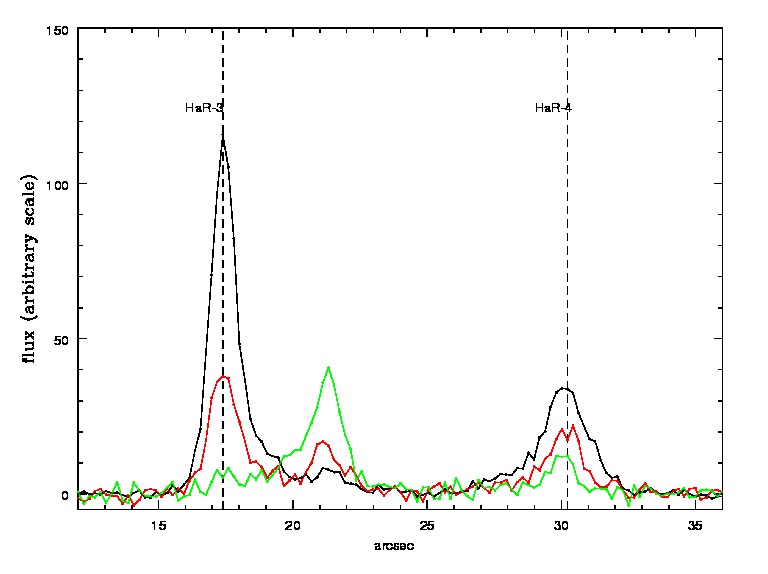}
\caption{Spatial distribution of \Ha(black), [NII]6584(red),
and the continuum(6654--6750\AA\ in observed frame; green). 
\Ha\ and [NII] include continuum flux.
[NII] is scaled $\times 2$, and the continuum is scaled $\times 5$ for
comparison with \Ha. The center of each aperture is shown as 
a vertical broken line. The origin of the coordinate is arbitrary.
(Top) HaR-1,2. HaR-2 shows several different sub-regions.
(bottom) HaR-3,4. The continuum peak at x=21 arcsec 
is background galaxy.
}
\label{fig:xprof}
\end{figure}

\begin{figure}
\includegraphics[scale=0.25,bb=0 0 768 574]{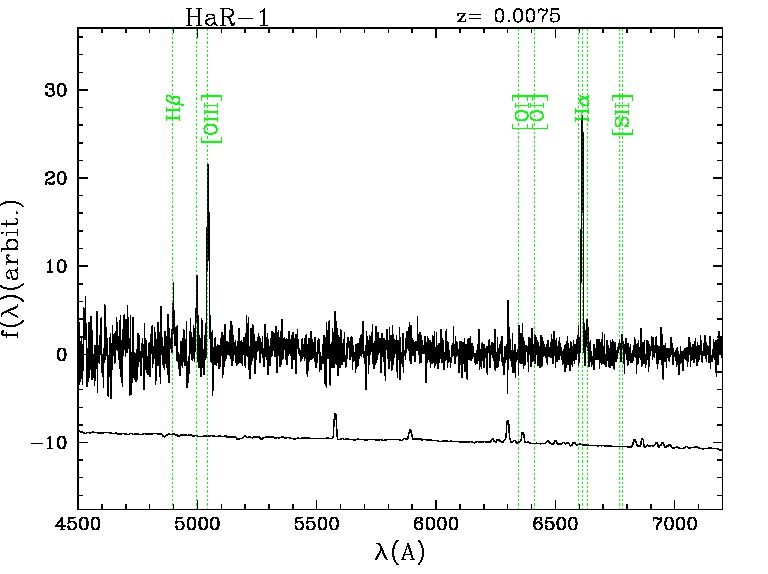}
\includegraphics[scale=0.25,bb=0 0 768 574]{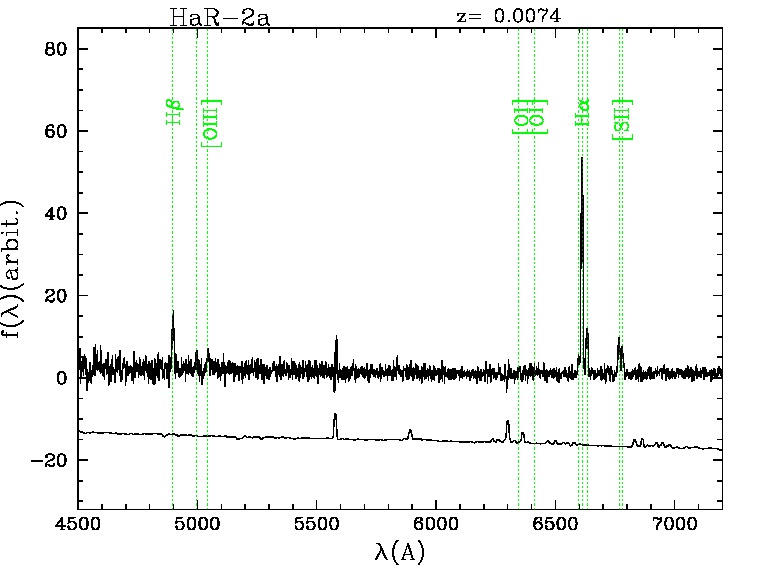}\\
\includegraphics[scale=0.25,bb=0 0 768 574]{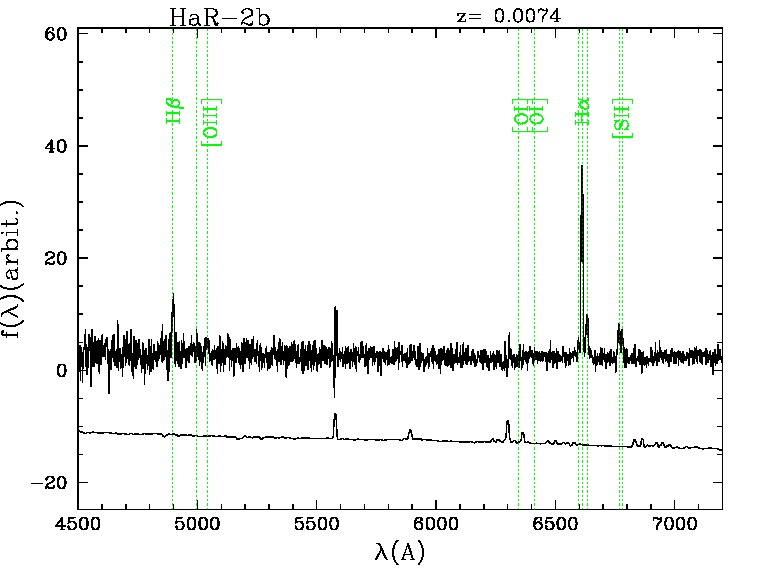}
\includegraphics[scale=0.25,bb=0 0 768 574]{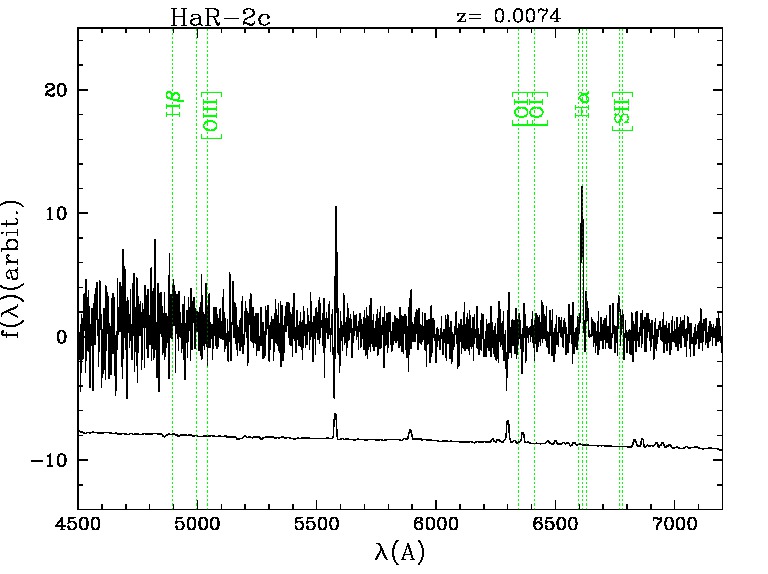}\\
\includegraphics[scale=0.25,bb=0 0 768 574]{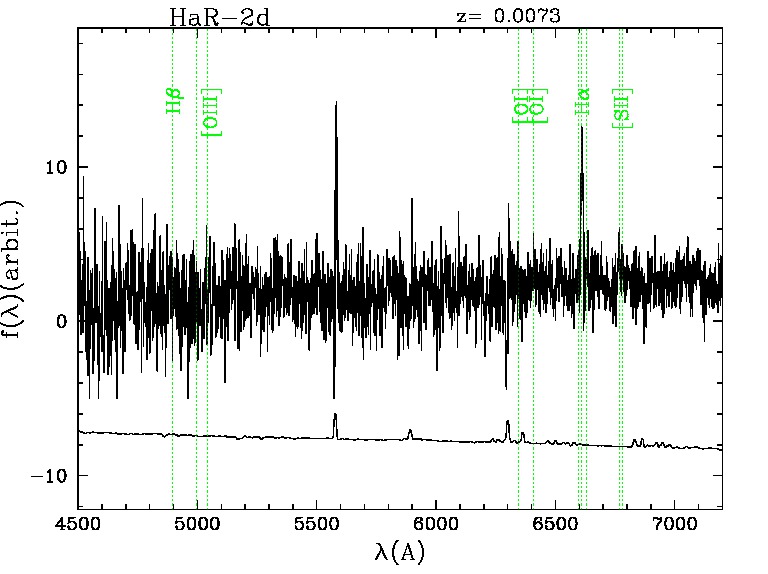}
\includegraphics[scale=0.25,bb=0 0 768 574]{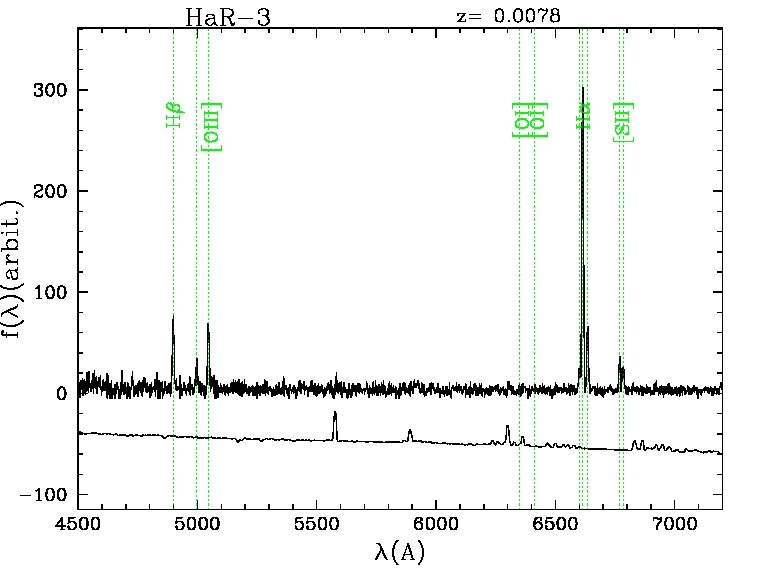}\\
\includegraphics[scale=0.25,bb=0 0 768 574]{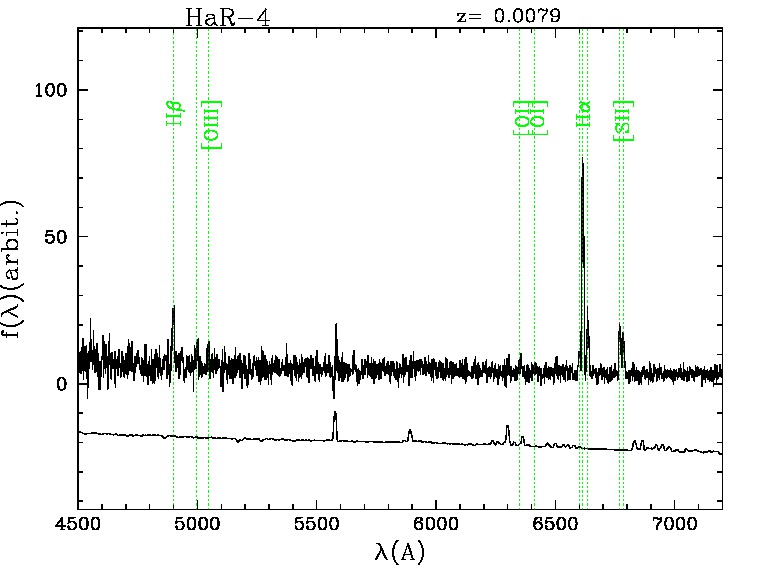}\\
\caption{The spectra of \Ha\ emitting regions.
From left to right and top to bottom, 
spectra of HaR-1(top-left), 2a, 2b, 2c, 2d,
3, and 4(bottom-left) are shown. The green lines indicate
the position of emission lines at the redshift in Table \ref{tab:redshift}.
The subtracted sky spectrum (scaled) is shown below each spectrum
for reference. The wavelength is in the rest frame.
}
\label{fig:spec}
\end{figure}

\begin{figure}
\includegraphics[scale=0.45,bb=0 0 769 575]{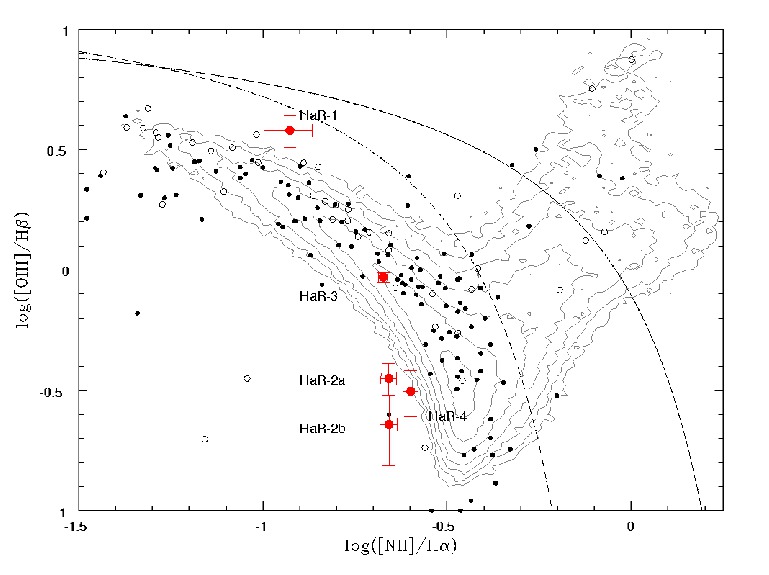}\\
\includegraphics[scale=0.45,bb=0 0 769 575]{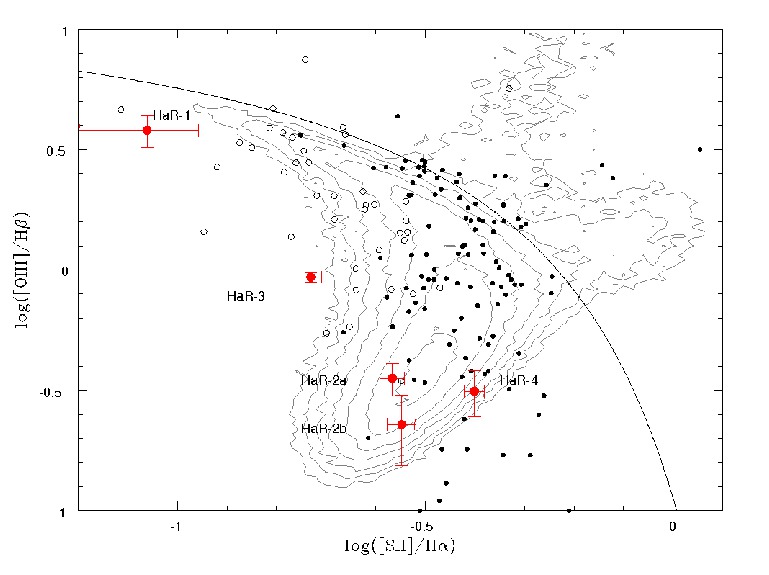}\\
\caption{Line ratio plot.
The large, filled circles with errorbars are the data of this study.
The solid and broken curves are the demarcation between \HII\ regions
and AGN by \citet{Kewley2001} and \citet{Kauffmann2003}, respectively.
Filled dots represent nearby field galaxies \citet{Jansen2000}, 
and open dots represent blue compact galaxies from \citet{Kong2002}.
The contour shows the distribution of SDSS galaxies
in the MPA-JHU DR7 release of spectrum measurements.
The data whose S/N of emission lines are larger than three are used,
and the number of galaxies are $4.0\times 10^5$ and $3.5\times 10^5$ for
[NII]/\Ha\ vs [OIII]/\Hb\ and
[SII]/\Ha\ vs [OIII]/\Hb, respectively.
The contour interval is a factor of $10^{1/4}$.
}
\label{fig:lineratios}
\end{figure}

\begin{figure}
\includegraphics[scale=0.22,bb=0 0 768 574]{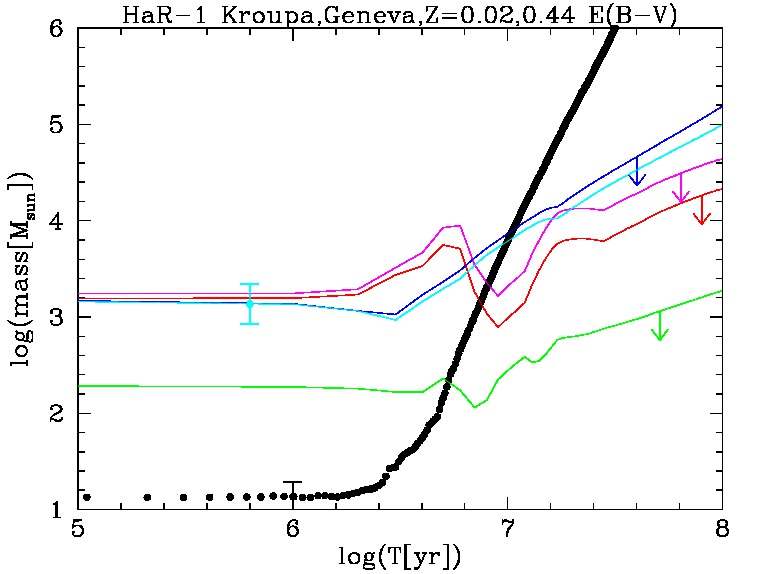}
\includegraphics[scale=0.22,bb=0 0 768 574]{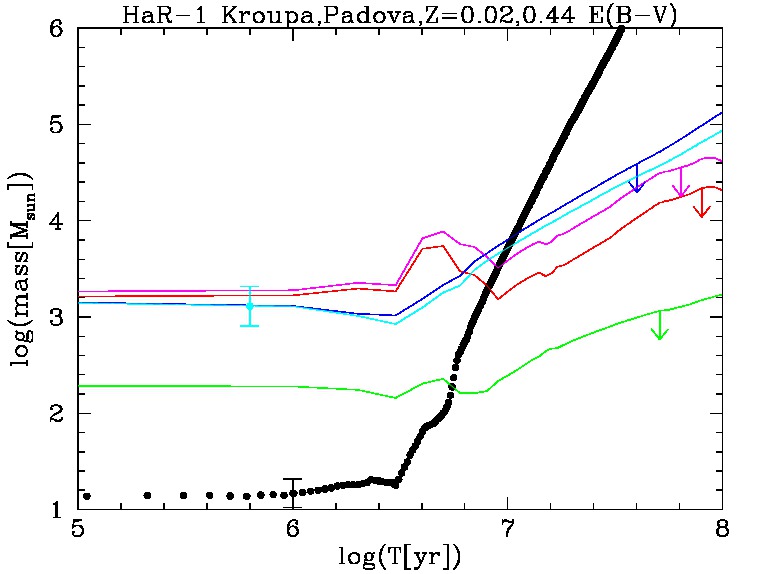}\\
\includegraphics[scale=0.22,bb=0 0 768 574]{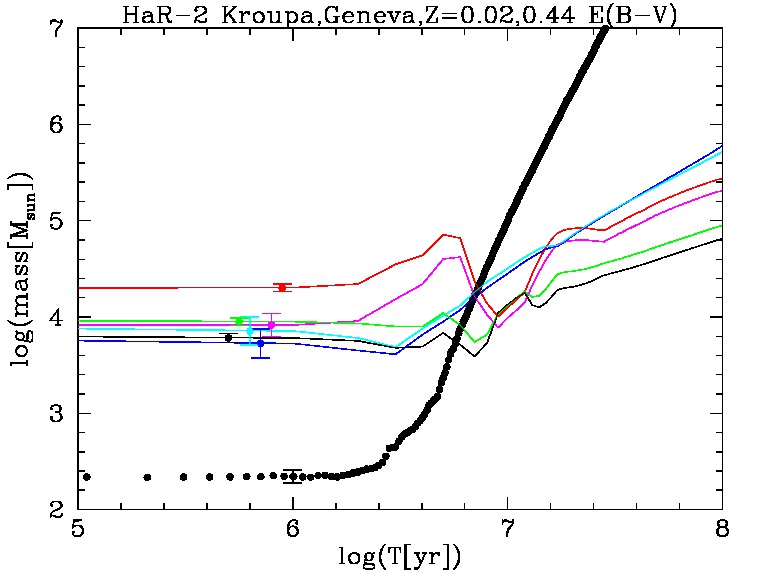}
\includegraphics[scale=0.22,bb=0 0 768 574]{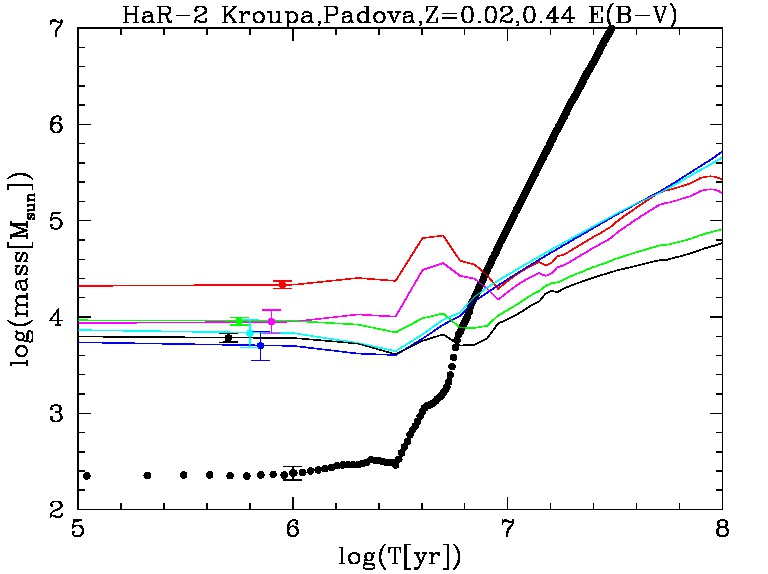}\\
\includegraphics[scale=0.22,bb=0 0 768 574]{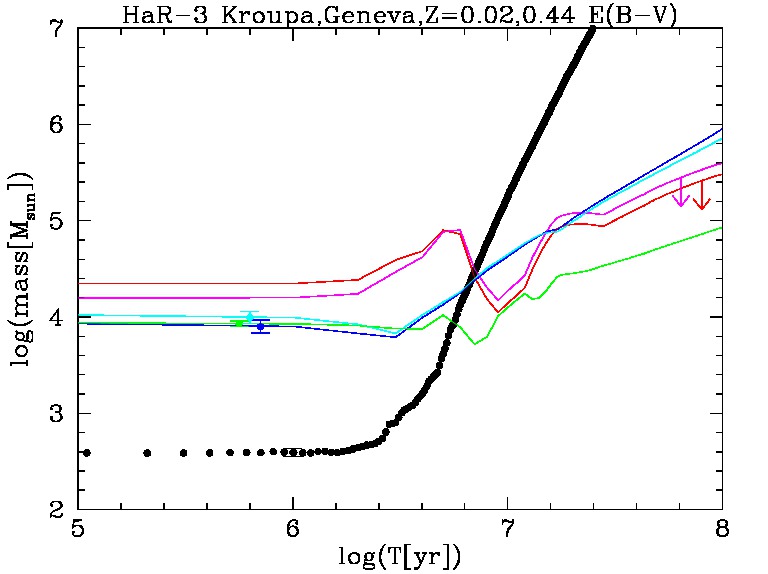}
\includegraphics[scale=0.22,bb=0 0 768 574]{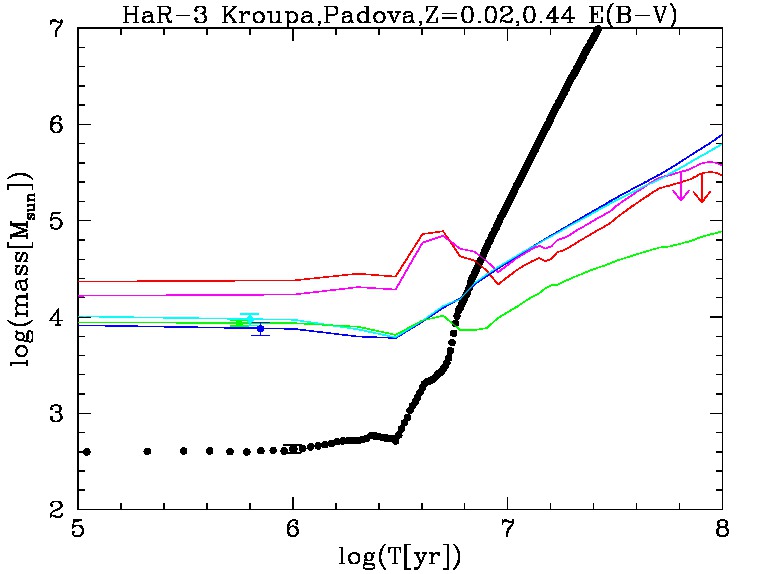}\\
\includegraphics[scale=0.22,bb=0 0 768 574]{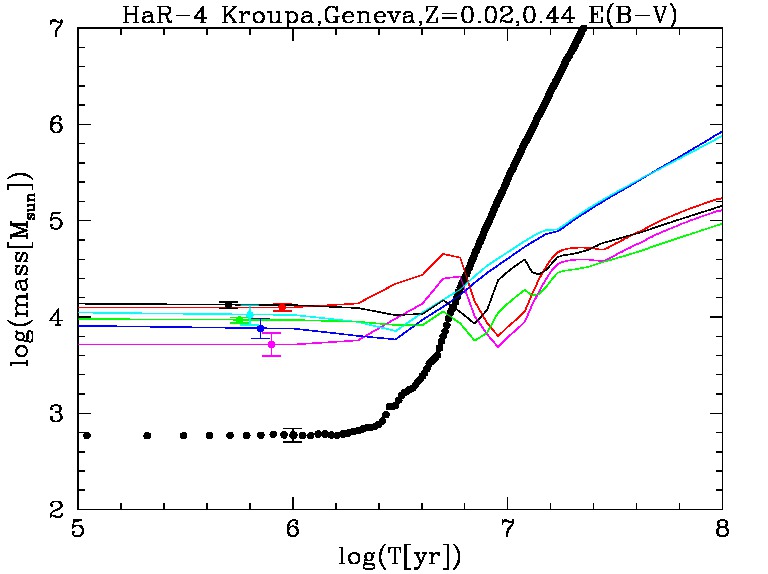}
\includegraphics[scale=0.22,bb=0 0 768 574]{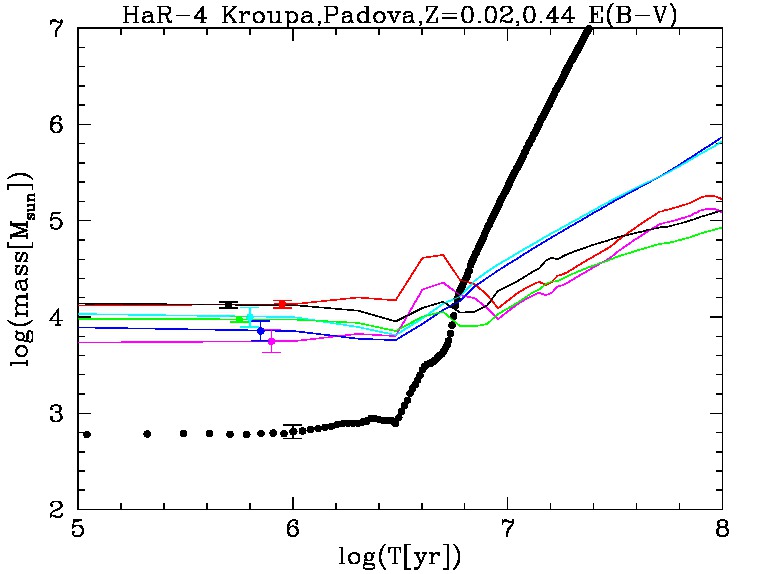}\\
\caption{Age and mass estimation using STARBURST99 model.
The combinations of the age(T) and mass to reproduce 
the observed \Ha\ luminosity (L(\Ha)) are shown as black dots.
Solid lines correspond to the combination to reproduce 
the observed magnitudes in 
FUV(blue), NUV(cyan), i(green), KPNO ha(black),
3.6$\mu$m(red), and 4.5$\mu$m(red). 
Detailed explanation is given in Appendix \label{sec:agemassdetail}.
Those on the left are Geneva track, and on the right are Padova track.
The arrows represent that the line indicate the upper limit.
The errorbars are shown around log(T)=6.
The error includes photometric error and error from E(B-V) estimation,
and is the same at all T. 
}
\label{fig:SB99}
\end{figure}

\begin{figure}
\includegraphics[scale=0.5,bb=0 0 892 892]{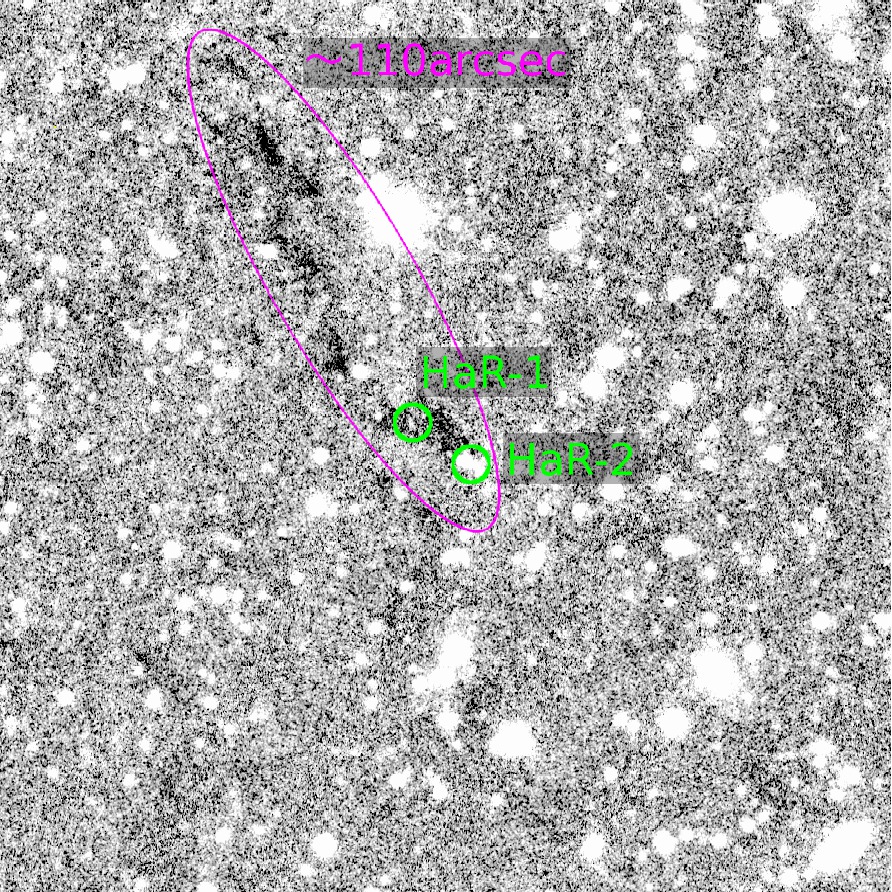}
\caption{
Wider view around HaR-1 and 2 in V-band with higher contrast
and grayscale inverted from Figure \ref{fig:stamps12}.
Magenta ellipse indicates a winding dark filament.
HaR-2 lay at the tip of the filament, and the length is about 
110 arcsec($\sim$9 kpc).
}
\label{fig:DC}
\end{figure}

\clearpage

\begin{table}
\begin{tabular}{|c|c|c|c|c|c|c|}
\hline
name & RA(J2000) & Dec(J2000)      \\
\hline 
HaR-1  & 12:26:14.63 & +12:51:46.4 \\
HaR-2a & 12:26:13.94 & +12:51:39.2 \\
HaR-2b & 12:26:13.87 & +12:51:38.6 \\
HaR-2c & 12:26:13.77 & +12:51:37.9 \\
HaR-2d & 12:26:13.68 & +12:51:36.8 \\
HaR-3  & 12:26:13.80 & +12:43:07.6 \\
HaR-4  & 12:26:13.09 & +12:43:00.4 \\
\hline
\end{tabular}
\caption{Position of targets.}
\label{tab:pos}
\end{table}

\begin{table}
{\small
\begin{tabular}{|c|c|c|c|c|c|c|c|}
\hline
name   & FUV & NUV & V & R & i & 3.6$\mu$m & 4.5$\mu$m \\
\hline 
HaR-1  & $>$24.4\tablenotemark{a} &24.7  $\pm$ 0.2 &$>$26.43\tablenotemark{b}&25.73$\pm$0.09&$>$26.3\tablenotemark{b}
                                                                                        &$>$24.3\tablenotemark{b}&$>$24.1\tablenotemark{b}\\        
HaR-2ab& ...\tablenotemark{c}&...\tablenotemark{c}                         &22.76$\pm$0.04&22.59$\pm$0.04&22.72$\pm$0.04 &22.1$\pm$0.1&23.0$\pm$0.3\\
HaR-2d & ...\tablenotemark{c}&...\tablenotemark{c}                          &22.86$\pm$0.06&22.68$\pm$0.04&22.52$\pm$0.04 &22.2$\pm$0.2&23.3$\pm$0.4\\
HaR-2  & 21.88 $\pm$ 0.09&21.85 $\pm$ 0.04&22.06$\pm$0.07&21.88$\pm$0.04&21.86$\pm$0.05 &21.5        & 22.4       \\
HaR-3  & 21.80 $\pm$ 0.09&21.86 $\pm$ 0.04&22.31$\pm$0.05\tablenotemark{d}&21.91$\pm$0.04\tablenotemark{d}&22.00$\pm$0.06\tablenotemark{d}
                                                                                        &$>$21.4\tablenotemark{e}&$>$21.7\tablenotemark{e}\\
HaR-4  & 21.24 $\pm$ 0.07&21.18 $\pm$ 0.03&21.53$\pm$0.04&21.08$\pm$0.04&21.76$\pm$0.04 &22.0$\pm$0.1&22.9$\pm$0.3\\
\hline
\end{tabular}
}
\tablenotetext{a}{No object shape, but has a positive flux in an aperture}
\tablenotetext{b}{1 sigma limiting magnitude}
\tablenotetext{c}{Not resolved}
\tablenotetext{d}{$\phi=2$ arcsec aperture photometry}
\tablenotetext{e}{twice of semicircular aperture flux}
\caption{Broadband Magnitudes after Galactic extinction correction}
\label{tab:photom}
\end{table}

\begin{table}
\begin{tabular}{|c|c|c|c|c|}
\hline
name   & NA503        & NA659        & ha & ha4\\
\hline 
HaR-1  &24.14$\pm$0.05&23.51$\pm$0.05&...\tablenotemark{a} & 23.22$\pm$0.06\\
HaR-2a &23.74$\pm$0.05&...\tablenotemark{b}           &23.86$\pm$0.14 &...\tablenotemark{b}\\
HaR-2b &23.48$\pm$0.06&...\tablenotemark{b}           &23.25$\pm$0.10 &...\tablenotemark{b}\\
HaR-2ab&...           &21.06$\pm$0.04&22.76$\pm$0.08 &21.09$\pm$0.02\\ 
HaR-2c &24.34$\pm$0.10&...\tablenotemark{a}           &...\tablenotemark{a}            &...\tablenotemark{a}\\
HaR-2d &23.83$\pm$0.07&22.45$\pm$0.04&23.52$\pm$0.10 & 21.89$\pm$0.05\\ 
HaR-2  &22.01$\pm$0.05&20.80$\pm$0.04&22.32$\pm$0.07 & 20.67$\pm$0.02\\
HaR-3  &22.20$\pm$0.04&20.19$\pm$0.04&...\tablenotemark{c} & 20.19$\pm$0.01\\
HaR-4  &21.72$\pm$0.04&19.87$\pm$0.04&21.39$\pm$0.04 & 19.80$\pm$0.01\\
\hline
\end{tabular}
\tablenotetext{a}{Non-detection}
\tablenotetext{b}{Not resolved}
\tablenotetext{c}{Heavily blended with background galaxy}
\caption{Narrowband Magnitudes after Galactic extinction correction}
\label{tab:photom2}
\end{table}

\begin{table}
\begin{tabular}{|c|c|c|c|}
\hline
name & z & v [km s$^{-1}$] & d[kpc]\\
\hline 
HaR-1  & 0.00749 & 2.25$\times 10^3$&67.1\\
HaR-2a & 0.00737 & 2.21$\times 10^3$&66.2\\
HaR-2b & 0.00737 & 2.21$\times 10^3$&66.1\\
HaR-2c & 0.00736 & 2.21$\times 10^3$&66.0\\
HaR-2d & 0.00731 & 2.19$\times 10^3$&65.9\\
HaR-3  & 0.00780 & 2.34$\times 10^3$&35.9\\
HaR-4  & 0.00789 & 2.37$\times 10^3$&34.9\\
\hline
\end{tabular}
\caption{Recession velocity in the heliocentric reference frame,
and projected distance from NGC~4388}
\label{tab:redshift}
\end{table}

\begin{table}
\begin{tabular}{|c|c|c|c|c|c|}
\hline
name & [NII]6584/\Ha & [SII]6717,6731/\Ha & [SII]6717/[SII]6731 & [OIII]5007/\Hb & \Ha/\Hb \\
\hline 
HaR-1  & 0.12$\pm$0.02 & 0.09$\pm$0.02 & 0.88$\pm$0.49  & 3.79$\pm$0.57 & 5.4$\pm$0.8\tablenotemark{a}\\
HaR-2a & 0.22$\pm$0.01 & 0.27$\pm$0.02 & 1.45$\pm$0.17  & 0.35$\pm$0.05 & 4.0$\pm$0.3\tablenotemark{a}\\
HaR-2b & 0.22$\pm$0.01 & 0.28$\pm$0.02 & 1.34$\pm$0.17  & 0.23$\pm$0.07 & 3.7$\pm$0.4\tablenotemark{a}\\
HaR-2c & 0.22$\pm$0.04 & 0.35$\pm$0.05 & 2.39$\pm$0.92  & 0.07$\pm$0.25 & 6.1$\pm$3.0\tablenotemark{a}\\
HaR-2d & 0.17$\pm$0.06 & 0.37$\pm$0.08 & 1.56$\pm$0.67  & ...\tablenotemark{b} & $>$4.4\tablenotemark{a,c} \\
HaR-3  & 0.22$\pm$0.01 & 0.19$\pm$0.01 & 1.44$\pm$0.14  & 0.93$\pm$0.04 & 4.4$\pm$0.2\\
HaR-4  & 0.25$\pm$0.01 & 0.40$\pm$0.02 & 1.29$\pm$0.11  & 0.31$\pm$0.07 & 3.7$\pm$0.3\\
\hline
\end{tabular}
\caption{line ratios}
\label{tab:lineratio}
\tablenotetext{a}{including 13\% error due to possible differential atmospheric dispersion effect}
\tablenotetext{b}{[OIII] and \Hb\ non-detection}
\tablenotetext{c}{from upper limit of \Hb}
\end{table}

\begin{table}
\begin{tabular}{|c|c|c|c|c|}
\hline
name & m(NA659) & E(B-V) & log(f(\Ha) [ergs s$^{-1}$ cm$^{-2}$]) & log(L(\Ha) [ergs s$^{-1}$])\\ 
\hline 
HaR-1  & 23.51$\pm$0.05 & 0.59$^{+0.13}_{-0.15}$  & -16.63$^{+0.15}_{-0.13}$ & 35.89$^{+0.15}_{-0.13}$\\
HaR-2a & 21.06$\pm$0.04\tablenotemark{a} 
                       & 0.31$\pm$0.07           & -15.45$\pm$0.07        & 37.07$\pm$0.07\\
HaR-2b & ...           & 0.23$^{+0.10}_{-0.11}$  & ...                      & ...\\
HaR-2c & ...           & 0.3$^{+0.2}_{-0.3}$     & ...                      & ... \\
HaR-2d & 22.45$\pm$0.04\tablenotemark{b} 
                       & $>$0.4                  & $>$-16.53                & $>$35.99\\
HaR-2 &  ...           & ...                     & ...                      & $>$37.10\\
HaR-3  & 20.19$\pm$0.04 & 0.40$\pm$0.04           & -15.17$\pm 0.04$         & 37.35$\pm 0.04$ \\
HaR-4  & 19.87$\pm$0.04 & 0.23$^{+0.07}_{-0.08}$  & -14.99$^{+0.08}_{-0.07}$ & 37.53$^{+0.08}_{-0.07}$ \\
\hline
\end{tabular}
\caption{\Ha\ luminosity}
\label{tab:Haflux}
\tablenotetext{a}{HaR-2a magnitude and flux includes HaR-2b and part of 2c region.}
\tablenotetext{b}{HaR-2d magnitude and flux includes part of 2c region.}
\end{table}

\begin{table}
\begin{tabular}{|c|c|c|c|c|l|}
\hline
cluster & galaxy & proj. dist.[kpc] & log(mass[M$_{\odot}$])& log(age[yr]) & reference \\
\hline
Abell~3627 & 137-001 & 0--40   &  ...    & ...     & \citet{Sun2010}\\
Coma       & RB199   & 23--35  & 5.8--7.1 & $\sim$8 & \citet{Yoshida2008}\\
           & IC~4040  & 1.1--48 &  ...    & $\sim$8 & \citet{Yoshida2012}\\
           & NGC~4848 & 22--35  &  ...    & ...     & \citet{Fossati2012}\\
Virgo      & IC~3418  & 10--17  & 4.5--5.6& 8.2--9.1& \citet{Fumagalli2011}\\
           &         & $\sim$14&  4      & 6.8(/7.2/7.3) & \citet{Ohyama2013}\\
           & NGC~4330 & $<$9    &3.3--6.3 & $(<7)$--8.5     & \citet{Abramson2011}\\
           & VCC~1249 & 5--12   &3.8--4.3 & 6.6--7.3& \citet{ArrigoniBattaia2012}\\
           & NGC~4388 & 66      & $<3$    & $<7$    & this study (HaR-1)\\
           &         & 35,66   & 4--4.5  & 6.7--6.9& this study (HaR-2,3,4)\\
\hline
\end{tabular}
\caption{Star-forming regions in a ram-pressure stripped tail of galaxy}
\label{tab:SFregions}
\tablenotetext{a}{Projected distance from the parent galaxy}
\end{table}
\clearpage

\appendix

\section{Detail of Photometric Data}

\subsection{Suprime-Cam}

The Suprime-Cam data used in the study are summarized in Table
\ref{tab:spcam}.  The data are retrieved from
SMOKA\footnote{\url{http://smoka.nao.ac.jp/}}.

The data were reduced the same way as in \citet{Yoshida2002}.
As the i-band data were obtained with FDCCDs,
crosstalk corrections were applied \citep{Yagi2012}.
The World Coordinate System was calibrated using 
wcstools 3.8.3 \citep{Mink2002} and Guide Star Catalog 2.3 \citep{Lasker2008}.
Photometric zero points 
were calibrated using the method given in \citet{Yagi2013}. 
We adopted the stars in Sloan Digital Sky Survey III
Data Release 9 as standard objects,
and converted the SDSS magnitudes to the Suprime-Cam system.
The detail of the construction of the magnitude conversion 
was described in \citet{Yagi2013}.
The color conversion coefficients 
from SDSS to Suprime-Cam magnitudes are given in Table \ref{tab:colorcoeff}.
Stars of $19<r<21$ magnitude in SDSS3 DR9 were used for our calibration.
The K-correct \citep{Blanton2007} v4 offset 
for SDSS\footnote{\url{http://howdy.physics.nyu.edu/index.php/Kcorrect}} 
is applied; $m_{AB}-m_{SDSS}$=0.012, 0.010 and 0.028 for $g$, $r$ and $i$, 
respectively.

In Figures \ref{fig:stamps12} and \ref{fig:stamps34}, 
net-\Ha\ images (NB659-R) are 
obtained by subtracting the scaled R-band image from the 
NA659-band image. As the seeing size was worse in R-band,
the NA659 image was convolved with a Gaussian (FWHM=0.95 arcsec)
to match the size of the point spread function (PSF).

We used SExtractor \citep{Bertin1996} for the photometry
of the Suprime-Cam data.
The Galactic extinction was corrected 
with the values in NED
based on \citet{Schlegel1998} and \citet{Schlafly2011};
$(A_V, A_R, A_i, A_{NA503}, A_{NA659})$
=(0.08, 0.06, 0.05, 0.09, 0.06)
and (0.08, 0.06, 0.05, 0.10, 0.07)
for  HaR-1,2 field and HaR-3,4 field, respectively.
Their magnitudes (MAG\_AUTO) after the Galactic extinction correction
are given in Tables \ref{tab:photom} and \ref{tab:photom2}.

\subsection{KPNO-4m}

NOAO science archive%
\footnote{\url{http://portal-nvo.noao.edu/}}
provides calibrated data around the region 
that were taken on 2007-04-22 and 2007-04-23
with NOAO MOSAIC 1 on the KPNO 4m Mayall telescope.
The data were already used in \citet{Kenney2008}.
We retrieved the data in rest-\Ha (ha; center=6566\AA, FWHM=80\AA\ at F=3.1),
\Ha\ around the Virgo redshift (ha4; center=6610.1\AA, FWHM=80\AA\ at F=3.1)
, and R-band (center=6622\AA, FWHM=1490\AA\ at F=3.1).
The filter characteristics are given on the KPNO web site%
\footnote{\url{http://www.noao.edu/kpno/mosaic/filters/}},
and the center and FWHM were calculated from the 
transmission data, and 15\AA\ shift of R-band at F=3.1.

SExtractor \citep{Bertin1996} was used for photometry.
The photometric zero-points were given in the FITS header.
We adopted the Galactic extinction of 0.06 for HaR-1 and HaR-2, 
and 0.07 for HaR-3 and HaR-4.
The AB-Vega offset of 0.25 mag was also applied.
The results are shown in Table \ref{tab:photom2}.

\subsection{GALEX}

In the GALEX archive%
\footnote{\url{http://galex.stsci.edu/GR6/}}, 
several coadded tiles covered the targets.
We retrieved the coadded FITS images listed in Table \ref{tab:GALEXtile},
and measured the flux in a circular aperture
in FUV (1344--1786\AA)  and NUV (1771--2831\AA) bands \citep{Morrissey2007}.
As the FWHM of the PSF of GALEX was large, 
we adopted the aperture size of 6 arcsec radius.
The files presented as 
{\it *\_skybg.fits} of each tile was used for 
the background correction.
The background subtracted aperture fluxes in the tiles 
were averaged with the weight of exposure time,
and 
converted to AB magnitude 
using the zero point given in \citet{Morrissey2007}.
We adopted the Galactic extinction correction in GALEX bands
by \citet{Wyder2007} as
$A_{FUV}=8.24 E(B-V)$, and $A_{NUV}=8.24 E(B-V) -0.67 E(B-V)^2$.
The results were $A_{FUV},A_{NUV}$=(0.206,0.206), (0.230,0.231).
for HaR-1,2 field and HaR-3,4 field, respectively.
The magnitude errors were estimated following the GALEX webpage%
\footnote{\url{http://galexgi.gsfc.nasa.gov/docs/galex/FAQ/counts_background.html}}.
The results are given in Table \ref{tab:photom}.
In FUV data, HaR-1 shows no recognizable shape.
We therefore assumed the measured aperture flux as the upper limit.

\subsection{XMM-OM}

The target fields were also observed by XMM-Newton Optical Monitor
(XMM-OM), and UVW1 \citep[center=2905\AA, width=620\AA][]{Kuntz2008} 
magnitudes of 
HaR-2,3, and 4 are available in the catalog in the archive.
The data were obtained from MAST%
\footnote{\url{http://archive.stsci.edu/}}.
The catalog magnitudes are shown in Table \ref{tab:XMM-OM}.
Although the error of magnitude was larger than that of GALEX,
the spatial resolution of XMM-OM was higher:
the FWHM of PSF was $\sim$5 arcsec and $\sim$2.3 arcsec
for GALEX NUV band and XMM-OM UVW1 band, respectively.
We checked that there is no apparent contamination 
from nearby objects around the targets in the 
UVW1 band of XMM-OM.

\subsection{SDSS}
HaR-4 was detected in SDSS3 DR9 as SDSS J122613.13+124300.4.
The SDSS magnitudes are shown in Table \ref{tab:SDSSmag}.
We corrected Galactic extinction and 
the offset of SDSS-magnitude from AB-magnitude
by the K-correct v4\citep{Blanton2007};
$m_{AB}-m_{SDSS}$=-0.036, 0.012, 0.010, 0.028, and 0.040 
for u, g, r, i, and z-band, respectively.
The i-band magnitude was comparable to our measurement in 
Suprime-Cam in Table \ref{tab:photom}.
Other three HaRs were not detected in SDSS.

We found that the model magnitude of HaR-4 
was $\sim$ 0.7 mag brighter in SDSS data release 7
\citep[DR7][]{DR7} than that in DR9.
The reason would be the difference of 
the adopted model profile of the object:
Petrosian radius was 7.359(DR7) arcsec vs 2.475(DR9) arcsec.
As seen in Figure \ref{fig:stamps34} bottom, 
whose image size is 20 arcsec square,
Petrosian radius of 7.3 arcsec was too large for HaR-4,
and we consider the magnitude in DR9 was appropriate.

The $\sim$ 0.7 mag difference was specific to HaR-4. 
Other brighter objects had comparable magnitude.
Our photometric calibration of Suprime-Cam data 
based on DR9 was not affected even if we used DR7.

\subsection{IRAC/Spitzer}
We retrieved calibrated data of InfraRed Array Camera (IRAC) of 
Spitzer Space Telescope in 3.6 $\mu$m (ch1) and 4.5 $\mu$m (ch2) channels 
from IRSA%
\footnote{\url{http://irsa.ipac.caltech.edu/}}.
Five images taken in February and March 2010 
covered our target fields.
Their unique observation identification number (AORKEY) 
and observation date are shown in Table \ref{tab:Spitzer}.
Exposure time was 93.6 and 96.8 seconds for ch1 and ch2, respectively.

We used the flux zero point given in the calibrated FITS headers
(1 DN pixel$^{-1}$ = 10$^6$Jy steradian$^{-1}$).
After background subtraction,
aperture photometry of 3 arcsec radius 
was performed for each target in each image.
Then, median flux was calculated for each target.

The error of each image was estimated from 
the median of the absolute deviation 
 of random aperture photometry of 3 arcsec radius.
AB magnitudes were calculated as
\begin{equation}
{\rm AB}=23.90-2.5\log(S[\mu {\rm Jy}]).
\end{equation}
The result is shown in Table \ref{tab:photom}.

HaR-1 was not recognized in the IRAC images, and the aperture flux was 
negative in some of the images.
The upper limit of aperture flux in 3 arcsec radius in ch1
(0.7$\mu$Jy) was therefore used.
HaR-3 was heavily blended with the neighbor galaxy.
We tried to measure the flux in a semicircular aperture 
to mask the side of the neighbor galaxy, and double the flux.
The value should still suffer from the envelope of the neighbor galaxy,
and we regard it as the upper limit of the flux.

\begin{table}
\begin{tabular}{|c|c|c|}
\hline
filter & DATE-OBS(UTC) & exptime \\
\hline
R & 2002-06-05 & 6min $\times$ 6\\
  & 2002-06-06 & 6min $\times$ 12\\
\hline 
NA659 & 2001-04-25 & 5.5min\\
      & 2001-04-26 & 20 min $\times$ 5\\
\hline 
V & 2001-03-24 & 5min $\times$ 3\\
\hline
NA503 & 2001-04-24 & 20min $\times$ 3 \\
\hline
i & 2010-04-12 & 2.8min $\times$ 5 + 1.4min\\
\hline
\end{tabular}
\caption{Archive data of Suprime-Cam used in this study}
\label{tab:spcam}
\end{table}

\begin{table}
\begin{tabular}{|c|c|c|c|c|c|c|c|c|c|c|}
\hline
SDSS-Suprime& CCD &SDSS& range &$c_0$&$c_1$&$c_2$ &$c_3$&$c_4$\\
\hline
$r-NA659$ & MIT & $r-i$ & 0$<r-i<$0.8    & -0.037 & 0.701 & -0.508 & 0.278 & ...\\
\hline
$g-NA503$ & MIT & $g-r$ & 0$<g-r<$0.6    & -0.046 & -0.228 & -0.101 & ... & ...\\
\hline
$g-V$     & SITe& $g-r$ & -0.4$<g-r<$0.8 & 0.039 & 0.574 & -0.030 & 0.271 & -0.221 \\
\hline
$i-i$     & HPK & $r-i$ & -0.4$<r-i<$1.6 & -0.006 & 0.089 & 0.019 & -0.015 & ...\\
\hline
\end{tabular}
\caption{The coefficients of color conversion}
\label{tab:colorcoeff}
\end{table}

\begin{table}[ht]
\begin{tabular}{|c|c|c|c|}
\hline
Tile & EXPTIME(FUV) & EXPTIME(NUV) & OBS-DATE\\
\hline
NGA\_Virgo\_MOS10    &1590.25& 3128.45 &2004-03-11\\
Virgo\_Epoque\_MOS01 &482.15 &15699.9  &2006-03-20\\
Virgo\_Epoque\_MOS08 &504.05 &16419.85 &2006-03-20\\
GI5\_057013\_NGC4388 &2538.  & 4993.8  &2009-05-07\\
\hline
\end{tabular}
\caption{GALEX tiles and exposure time}
\label{tab:GALEXtile}
\end{table}

\begin{table}[ht]
\begin{tabular}{|c|c|c|c|}
\hline
Name& XMM-OM dataset & UVW1 mag&DATE-OBS\\
& & (ABmag)& \\
\hline
HaR-1 &0108260201& ...           &2002-07-02\\ 
HaR-2 &0108260201& 21.9 $\pm$0.4 &2002-07-02\\
HaR-3 &0110930301& 21.5$\pm$0.5  &2002-07-07\\
HaR-4 &0110930301& 21.1$\pm$0.3  &2002-07-07\\
      &0110930701& 21.2$\pm$0.5  &2002-12-12\\
\hline
\end{tabular}
\caption{XMM-OM}
\label{tab:XMM-OM}
\end{table}

\begin{table}[ht]
\begin{tabular}{|c|c|c|c|c|c|}
\hline
object & u & g & r & i & z\\
\hline
HaR-4 & 22.4$\pm$0.5 & 21.31$\pm$0.08 & 21.38$\pm$0.12 & 21.55$\pm$0.16 &
21.6$\pm$0.6 \\
\hline
\end{tabular}
\caption{AB magnitudes of HaR-4 from SDSS-DR9}
\label{tab:SDSSmag}
\end{table}

\begin{table}[ht]
\begin{tabular}{|c|c|}
\hline
AORKEY & DATE\_OBS \\
\hline
35323392 &2010-03-11\\ 
35324160 &2010-03-10\\
35325184 &2010-03-08\\
35325696 &2010-03-03\\
35326208 &2010-02-24\\
\hline
\end{tabular}
\caption{Spitzer}
\label{tab:Spitzer}
\end{table}

\section{Detail of the age and mass estimation}
\label{sec:agemassdetail}

For age and mass estimation of HaRs,
we used STARBURST99 v6.0.4\citep{Leitherer1999,Vazquez2005,Leitherer2010} model 
and calculate the magnitudes and \Ha\ luminosity.
We adopted the metallicity Z=0.02, 
the IMF by \citet{Kroupa2001},
an instantaneous burst of a fixed mass,
and both Geneva and Padova stellar evolutionary tracks.
We also tested Salpeter IMF
\citep{Salpeter1955}, but the result did not change significantly.
The spectra were redshifted to the velocity of HaR-2: z=0.0074.
We checked that the redshift difference of HaR-1,2(z=0.0074) and HaR-3,4(z=0.0078) 
made only $<0.001$ mag difference, and we neglected the effect.

The model magnitudes were calculated from the spectrum of each age
multiplied with system responses of GALEX,
KPNO mosaic, IRAC, 
and Suprime-Cam/Subaru.
The filter transmittances of GALEX were obtained from COSMOS filter
response page\footnote{\url{http://www.astro.caltech.edu/~capak/cosmos/filters/}},
which data was converted from \citet{Morrissey2005}.
The filter transmittances of IRAC were taken from Spitzer Document page 
at IRSA%
\footnote{\url{http://irsa.ipac.caltech.edu/data/SPITZER/docs/irac/calibrationfiles/spectralresponse/}}.
The filter transmittances of Suprime-Cam were constructed as the product of 
the quantum efficiency of CCDs%
\footnote{\url{http://www.naoj.org/Observing/Instruments/SCam/ccd.html}},
the transmittance of the primary focus corrector%
\footnote{\url{http://www.naoj.org/Observing/Telescope/Parameters/PFU/}},
the reflectivity of the primary mirror%
\footnote{\url{http://www.naoj.org/Observing/Telescope/Parameters/Reflectivity/}},
the atmospheric extinction model used in \citet{Yagi2013},
and 
the filter responses in Subaru WWW for i-band
\footnote{\url{http://www.naoj.org/Observing/Instruments/SCam/sensitivity.html}}.
The simulated filter responses of KPNO mosaic Ha at F/3.1 focus was taken 
from the NOAO web page%
\footnote{\url{http://www.noao.edu/kpno/mosaic/filters/}}.
Note that we did not use V, R, NA503, and NA659 of Suprime-Cam and 
ha4 of KPNO in this fitting,
since they would be contaminated by strong nebular emission lines.
As the transmittance of Suprime-Cam and KPNO mosaic 
was measured in the air wavelength, the model spectra were converted to 
the air wavelength using the index of \citet{Ciddor1996},
when these magnitudes were calculated. 
The internal extinction was calculated as
$E_{\rm star}(B-V)=0.44 E_{\rm gas}(B-V)$ \citep{Calzetti2000},
where $E_{\rm gas}(B-V)$ are given in Table \ref{tab:Haflux}.
The coefficient $R_{\rm band} = A_{\rm band}/E_{\rm star}(B-V)$ 
we adopted are 
(FUV,NUV,V,R,i,3.6$\mu$,4.5$\mu$)=(8.24,8.24,3.1,2.5,1.9,0.203,0.156),
respectively.

\begin{figure}[htb]
\includegraphics[scale=0.25,bb=0 0 768 574]{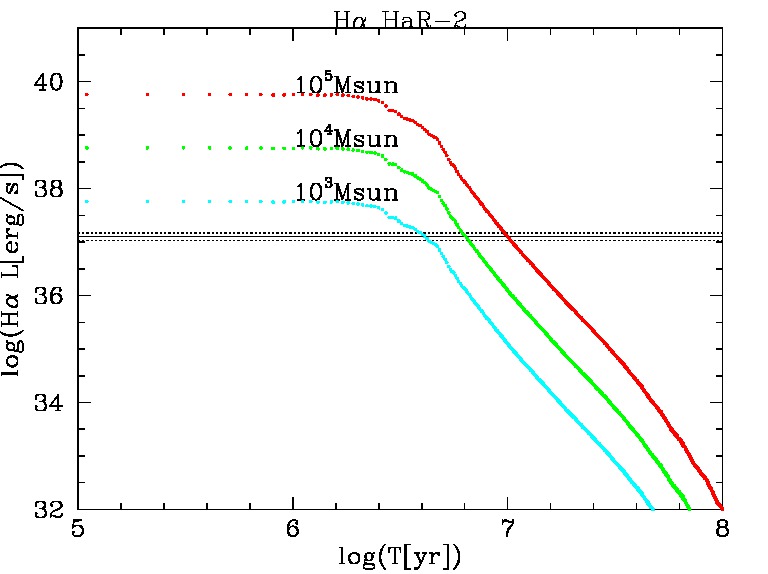}
\includegraphics[scale=0.25,bb=0 0 768 574]{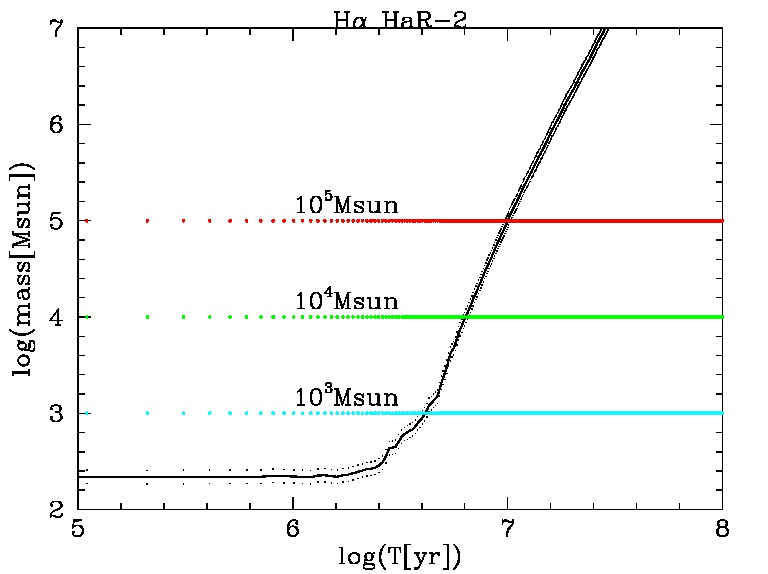}\\
\includegraphics[scale=0.25,bb=0 0 768 574]{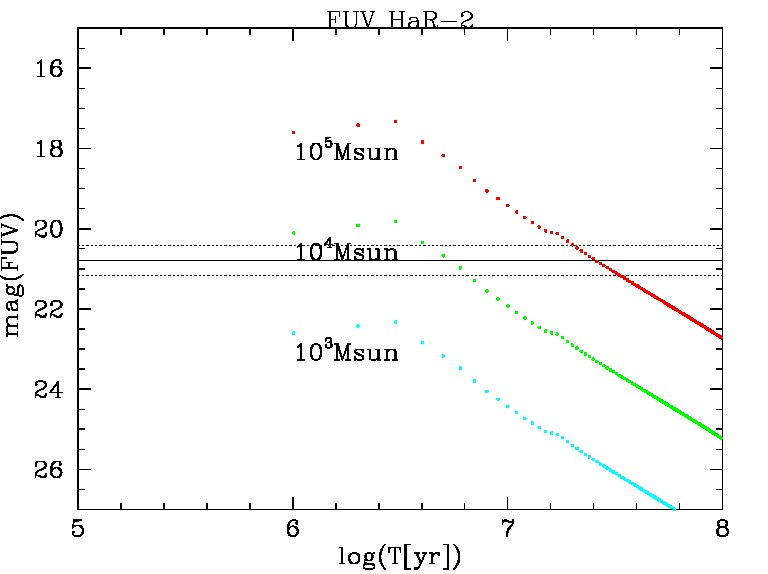}
\includegraphics[scale=0.25,bb=0 0 768 574]{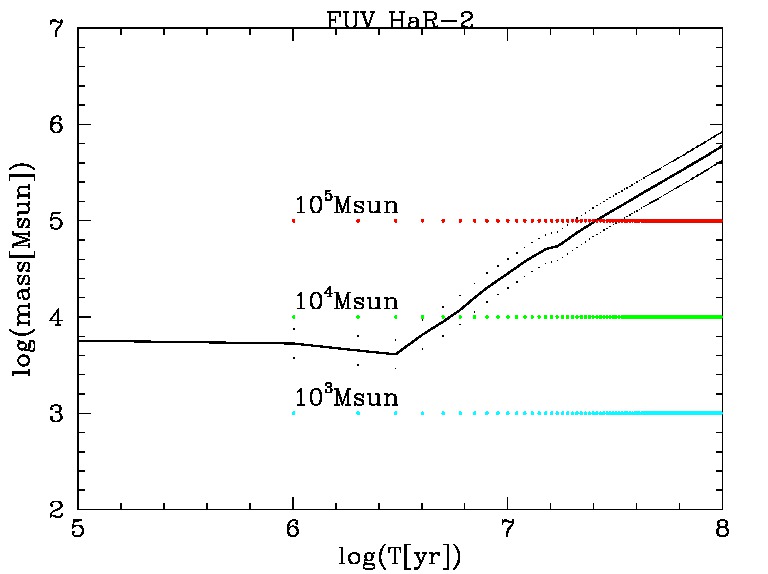}\\
\caption{
(top-left) Model \Ha\ luminosity of clouds from STARBURST99 calculation. 
Geneva model was adopted.
The horizontal lines show the observed luminosity (solid) and
its error (broken).
(too-right) Transformed from the left figure so that the ordinate to be 
log(mass). Data were shifted at each age
by $-\log({\rm L(H\alpha)/mass})$.
Now the observed data is a curve.
(bottom-left) Same as top-left,
but of FUV magnitude.
(bottom-right) Same as top-right,
but converted from FUV magnitude by multiplying by -2.5 
and shifting by $2.5~{\rm mag}_{FUV}+\log({\rm mass})$.
}
\label{fig:schematic}
\end{figure}

From the model calculation, 
we can plot the log of the \Ha\ luminosity (L(\Ha)) as a function
of the logarithm of the age T as Figure \ref{fig:schematic} top-left.
In the figure, observed 
L(\Ha) is shown as a solid 
horizontal line, and the 1-$\sigma$ of the L(\Ha) 
are shown as broken horizontal lines.
The crossing is the solution where the model and 
the observation are consistent.
By adding an offset of
$-\log({\rm L(H\alpha)/mass})$ of the model 
at each age, the figure is transformed 
so that the ordinate corresponds to the initial mass 
of the cluster 
(Figure \ref{fig:schematic} top-right).
In the figure, models
are horizontal (since the initial mass is time-invariant),
and observed value is a curve.
The error of the curve comes from the observed value, 
including the uncertainty of 
the internal extinction correction, and the amount of the error is 
the same at each age. No error of age at each point.

Similar transformation is applicable to the plots of magnitudes.
For example, the FUV magnitude is plotted as 
Figure \ref{fig:schematic} bottom-left.
It is converted to Figure \ref{fig:schematic} bottom-right
by multiplying by -2.5 
and adding an offset of $-\log({\rm l(FUV)/mass})$ of the model 
at each age, where log(l(FUV))=-2.5 mag$_{FUV}$.

Since the coordinates of the transformed figures (right panels of Figure 
\ref{fig:schematic}) are the same,
we can overplot them in a single figure, with magnitudes in other bands.
Figures \ref{fig:SB99} were thus constructed.
And the crossing of these curves is the solution where
the models and the observations are consistent.

\begin{thebibliography}{}

\bibitem[Abazajian et al.(2009)]{DR7}
Abazajian, K. N. et al. 2009, \apjs, 182, 543

\bibitem[Abramson et al.(2011)]{Abramson2011}
Abramson, A. 2011, \aj, 141, 164

\bibitem[Ahn et al.(2012)]{DR9}
Ahn, et al. 2012, \apjs, 203, 21

\bibitem[Arnabildi et al.(2003)]{Arnaboldi2003}
Arnaboldi, M. et al. 2003, \aj, 125, 514

\bibitem[Arrigoni Battaia et al.(2012)]{ArrigoniBattaia2012}
Arrigoni Battaia, F. et al. 2012, \aap, 543, A112

\bibitem[Bertin \& Arnouts(1996)]{Bertin1996}
Bertin, E., Arnouts, S.
1996, \aaps, 317, 393

\bibitem[Blanton \& Roweis(2007)]{Blanton2007}
Blanton, M.R., Roweis, S. 2007, \aj, 133, 734

\bibitem[Boissier et al.(2012)]{Boissier2012}
Boissier, S. et al. 2012, \aap, 545A, 142

\bibitem[Boselli \& Gavazzi(2006)]{Boselli2006}
Boselli, A., Gavazzi, G. 2006, \pasp, 118, 517

\bibitem[Calzetti et al.(2000)]{Calzetti2000}
Calzetti, D. et al. 2000, \apj, 533, 682

\bibitem[Calzetti et al.(1994)]{Calzetti1994}
Calzetti, D., Kinney, A.L., Storchi-Bergmann, T. 1994, \apj, 429, 582

\bibitem[Chung et al.(2007)]{Chung2007}
Chung, A., van Gorkom, J.H., Kenney, J.D.P., Vollmer, B. 2007, \apj, 659, L115

\bibitem[Ciddor(1996)]{Ciddor1996} 
Ciddor, P.E. 1996, Applied Optics, 35, 1566

\bibitem[Conselice, Gallagher\&Wyse(2001)]{Conselice2001}
Conselice, C.J., Gallagher, J.S. III, Wyse, R.F.G. 2001, \apj, 559, 791

\bibitem[Cortese et al.(2003)]{Cortese2003}
Cortese, L. et al. 2003, \aap, 401, 471 

\bibitem[Cortese et al.(2004)]{Cortese2004}
Cortese, L., Gavazzi, G., Boselli, A., Iglesias-Paramo, J. 2004, \aap, 416, 119

\bibitem[Cortese et al.(2007)]{Cortese2007}
Cortese, L. et al. 2007, \mnras, 376, 157


\bibitem[Cortese et al.(2010a)]{Cortese2010a}
Cortese, L. et al. 2010a, \aap, 518, L49

\bibitem[Cortese et al.(2010b)]{Cortese2010b}
Cortese, L. et al. 2010b, \aap, 518, L63

\bibitem[Crowl et al.(2005)]{Crowl2005}
Crowl, H.H., Kenney, J.D.P., van Gorkom, J.H., Vollmer, B. \aj, 130, 65

\bibitem[Ebeling et al.(1998)]{Ebeling1998}
Ebeling, H. et al. 1998, \mnras, 301, 881

\bibitem[Elmegreen(1989)]{Elmegreen1989} 
Elmegreen, B.~G. 1989, \apj, 338, 178

\bibitem[Elmegreen\& Efremov(1997)]{Elmegreen1997} 
Elmegreen, B.~G., \& Efremov, Y.~N. 1997, \apj, 480, 235

\bibitem[Fossati et al.(2012)]{Fossati2012}
Fossati, M., Gavazzi, G., Boselli, A., Fumagalli, M. 2012, \aap, 544, A128

\bibitem[Fujita \& Nagashima(1999)]{Fujita1999}
Fujita, Y., Nagashima, M. 1999, \apj, 516, 619
 
\bibitem[Fujita et al.(2006)]{Fujita2006}
Fujita, Y., Sarazin, C.L., Sivakoff, G.R. 2006, \pasj, 58, 131

\bibitem[Fumagalli et al.(2011)]{Fumagalli2011}
Fumagalli, M., Gavazzi, G., Scaramella, S., Franzetti, P. 2011,
\aap, 528, A46

\bibitem[Gavazzi et al.(2000)]{Gavazzi2000}
Gavazzi, G., et al. 2000, \aap, 361, 1

\bibitem[Gavazzi et al.(2001)]{Gavazzi2001}
Gavazzi, G., et al. 2001, \apjl, 563, L23

\bibitem[Gerhard et al.(2002)]{Gerhard2002}
Gerhard, O., Arnaboldi, M., Freeman, K.C., Okamura, S. 2002, \apj, 580, L121

\bibitem[Grevesse et al.(2010)]{Grevesse2010}
Grevesse, N., Asplund, M., Sauval, A.J., Scott, P. 2010, \apss, 328, 179

\bibitem[Giovanelli et al.(2007)]{Giovanelli2007}
Giovanelli, R. et al. 2007, \aj, 133, 2569

\bibitem[Gu et al.(2013)]{Gu2013}
Gu, L. et al. 2013, submitted to ApJL (paper I; arXiv:1308.5760)

\bibitem[Gunn \& Gott(1972)]{Gunn1972}
Gunn, J.E., Gott, J.R. III \apj, 176, 1

\bibitem[Hayashino et al.(2003)]{Hayashino2003}
Hayashino, T. et al. 2003, PNAOJ, 7, 33

\bibitem[Hester et al.(2010)]{Hester2010}
Hester, J.A. 2010, \apjl, 716, L14

\bibitem[Iwasawa et al.(2003)]{Iwasawa2003}
Iwasawa, K., Wilson, A.S, Fabian, A.C., Young, A.J. 2003, \mnras, 345, 369

\bibitem[Jansen et al.(2000)]{Jansen2000}
Jansen, R.A., Fabricant, D., Franx, M., Caldwell, N. 2000, \apjs, 126, 331

\bibitem[Kauffmann et al.(2003)]{Kauffmann2003}
Kauffmann, G. et al. 2003, \mnras, 346, 1055

\bibitem[Kashikawa et al.(2002)]{Kashikawa2002} 
Kashikawa, N. et al. 2002, \pasj, 54, 819

\bibitem[Kapferer et al.(2009)]{Kapferer2009}
Kapferer, W., Sluka, C., Schindler, S., Ferrari, C., Ziegler, B.
2009, \aap, 499, 87

\bibitem[Kenney et al.(2008)]{Kenney2008}
Kenney, J.D.P., Tal, T., Crowl, H.H., Feldmeier, J., Jacoby, G.H. 2008, \apjl, 687, L69

\bibitem[Kepferer et al.(2008)]{Kepferer2008}
Kapferer, W. et al. 2008, \aap, 499, 87

\bibitem[Kewley et al.(2001)]{Kewley2001}
Kewley, L.J,, Dopita, M.A., Sutherland, R.S.,
Heisler, C.A., Trevena, J. 2001, \apj, 556, 121

\bibitem[Kewley\&Dopita(2002)]{Kewley2002}
Kewley, L.J., Dopita, M.A. 2002, \apjs, 142, 35

\bibitem[Koda et al.(2012)]{Koda2012}
Koda, J., et al. 2012, \apj, 749, 20

\bibitem[Kong et al.(2002)]{Kong2002}
Kong, X., Cheng, F.Z., Weiss, A., Charlot, S. 2002, \aap, 396, 503


\bibitem[Kronberger et al.(2008)]{Kronberger2008}
Kronberger, T. et al. 2008, \aap, 481, 337

\bibitem[Kroupa(2001)]{Kroupa2001}
Kroupa, P. 2001, \mnras, 322, 231

\bibitem[Kuntz et al.(2008)]{Kuntz2008}
Kuntz, K.D., Harrus, I., McGlynn, T.A., Mushotzky, R.F., Snowden, S.L.
2008, \pasp, 120, 740

\bibitem[Lasker et al.(2008)]{Lasker2008}
Lasker B. et al. 2008, \aj, 136, 735

\bibitem[Leitherer et al.(1999)]{Leitherer1999}
Leitherer, C., et al. 1999, \apjs, 123, 3 

\bibitem[Leitherer et al.(2010)]{Leitherer2010}
Leitherer, C., et al. 2010, \apjs, 189, 309

\bibitem[Lu et al.(1993)]{Lu1993}
Lu, N.Y. et al. 1993, \apjs, 88, 383

\bibitem[Machacek et al.(2005)]{Machacek2005}
Machacek, M. et al. 2005, \apj, 630, 280

\bibitem[Machacek et al.(2006)]{Machacek2006}
Machacek, M. et al. 2006, \apj, 644, 145

\bibitem[Mink(2002)]{Mink2002}
Mink, D.J. 2002, in ASP Conf. Proc. 281. ADASS XI,
eds. D.A. Bohlender, D. Durand, \& T.H. Handley (San Francisco: ASP), 169

\bibitem[Miyazaki et al.(2002)]{Miyazaki2002} 
Miyazaki, S., et al. 2002, \pasj, 54, 833

\bibitem[Morrissey et al.(2005)]{Morrissey2005}
Morrissey, P. et al. 2005, \apjl, 619, L7

\bibitem[Morrissey et al.(2007)]{Morrissey2007}
Morrissey, P. et al. 2007, \apjs, 173, 682

\bibitem[Murray et al.(1993)]{Murray1993} 
Murray, S.~D., White, S.~D.~M., Blondin, J.~M., Lin, D.~N.~C. 
1993, \apj, 407, 588

\bibitem[O'Dell et al.(2013)]{ODell2013}
O'Dell, C.R., Ferland, G.J., Henney, W.J. Peimbert, M. 2013, \aj, 145, 93

\bibitem[Osterbrock\&Ferland(2006)]{Osterbrock2006}
Osterbrock, D.E., Ferland, G.J. 2006, 
``Astrophysics of gaseous nebulae and active galactic nuclei, 2nd. ed.'' 
(Sausalito, CA: University Science Books)

\bibitem[Ohyama \& Hota(2013)]{Ohyama2013}
Ohyama, Y., Hota, A. 2013, \apj, 767, L29

\bibitem[Okamura et al.(2002)]{Okamura2002}
Okamura, S. et al. 2002, \pasj, 54, 883


\bibitem[Oke \& Gunn(1983)]{Oke1983}
Oke, J.B., Gunn, J.R. 1983, \apj, 266, 713

\bibitem[Oosterloo \& van Gorkom (2005)]{Oosterloo2005}
Oosterloo, T., van Gorkom, J. 2005, \aap, 437, L19

\bibitem[Owen et al.(2006)]{Owen2006}
Owen, F.N. et al. 2006, \aj, 131, 1974
 
\bibitem[Pickles(1998)]{Pickles1998}
Pickles, A. 1998, \pasp, 110, 863

\bibitem[Roediger et al.(2006)]{Roediger2006}
Roediger, E., Br\"uggen, M., Hoeft, M. 2006, \mnras, 371, 609

\bibitem[Roediger\&Br\"uggen(2007)]{Roediger2007}
Roediger, E., Br\"uggen,, M. 2007, \mnras, 380, 1399

\bibitem[Roediger\&Br\"uggen(2008)]{Roediger2008}
Roediger, E., Br\"uggen, M. 2008, \mnras, 388, 486

\bibitem[Roediger(2009)]{Roediger2009}
Roediger, E. 2009, Astronomische Nachrichten, 330, 888

\bibitem[Salpeter(1955)]{Salpeter1955}
Salpeter, E.E. 1955, \apj, 121, 161

\bibitem[Schlegel et al.(1998)]{Schlegel1998}
Schlegel, D., Finkbeiner, D.P., and Davis, M. 1998, \apj, 500, 525

\bibitem[Schlalfy\&Finkbeiner(2011)]{Schlafly2011}
Schlafly, E.F., Finkbeiner, D.P. 2011, \apj, 737, 103

\bibitem[Sivanandam et al.(2010)]{Sivanandam2010}
Sivanandam, S, Rieke, M.J., Rieke, G.H. 2010, \apj, 711, 147

\bibitem[Smith et al.(2010)]{Smith2010}
Smith, R.J. et al. 2010, \mnras, 408, 1471

\bibitem[Sternberg et al(2003)Sternberg, Hoffmann, \&Pauldrach]{Sternberg2003}
Sternberg, A., Hoffmann, T.L., Pauldrach, A.W.A 2003, \apj, 599, 1333

\bibitem[Sun\&Vikhlinin(2005)]{Sun2005}
Sun, M., Vikhlinin, A. 2005, \apj, 621, 718

\bibitem[Sun et al.(2006)]{Sun2006}
Sun, M., Jones, C., Forman, W., Nulsen, P.E.J., Donahue, M.,
Voit, G.M. 2006, \apjl, 637, L81

\bibitem[Sun et al.(2007)]{Sun2007}
Sun, M., Donahue, M., Voit, G.M. 2007 \apj, 671, 190

\bibitem[Sun et al.(2010)]{Sun2010}
Sun, M. et al. 2010, \apj, 708, 946

\bibitem[Tonnesen\&Bryan(2012)]{Tonnesen2012}
Tonnesen, S., Bryan, G.L. 2012, \mnras, 422, 1609

\bibitem[van Dokkum(2001)]{vanDokkum2001}
van Dokkum, P.G. 2001, \pasp, 113, 1420

\bibitem[Vazquez\&Leitherer(2005)]{Vazquez2005}
Vazquez, V.A., Leitherer, C.2005, \apj, 621, 695

\bibitem[Vollmer\&Huchtmeier(2007)]{Vollmer2007}
Vollmer, B., Huchtmeier, W. 2007, \aap, 462, 93 

\bibitem[Vollmer et al.(2008)]{Vollmer2008}
Vollmer, B., Braine, J., Pappalardo, C., Hily-Blant, P. 2008, \aap, 491, 455

\bibitem[Vollmer (2009)]{Vollmer2009}
Vollmer, B. 2009, \aap, 502, 427

\bibitem[Vollmer et al.(2012)]{Vollmer2012}
Vollmer, B. et al. 2012, \aap, 537, A143


\bibitem[Wang et al.(2004)]{Wang2004}
Wang, Q.D., Owen, F., Ledlow, M. 2004, \apj, 611, 821

\bibitem[We\.zgowiec et al.(2011)]{Wezgowiec2011}
We\.zgowiec, M. et al. 2011, \aap, 531, A44

\bibitem[Wegner(2006)]{Wegner2006}
Wegner, W. 2006, \mnras, 371, 185

\bibitem[Wyder et al.(2007)]{Wyder2007}
Wyder, T.K. et al. \apjs, 173, 293

\bibitem[Yagi et al.(2007)]{Yagi2007}
Yagi, M., et al. 2007, \apj, 660, 1209

\bibitem[Yagi et al.(2010)]{Yagi2010}
Yagi, M., et al. 2010, \aj, 140, 1814

\bibitem[Yagi(2012)]{Yagi2012}
Yagi, M. 2012, \pasp, 124, 1347

\bibitem[Yagi et al.(2013)]{Yagi2013}
Yagi, M., et al. 2013, \pasj, 65, 22

\bibitem[Yamagami\&Fujita(2011)]{Yamagami2011}
Yamagami, T., Fujita, Y. 2011, \pasj, 63, 1165

\bibitem[Yoshida et al.(2002)]{Yoshida2002}
Yoshida, M., et al. 2002, \apj, 567, 118

\bibitem[Yoshida et al.(2004)]{Yoshida2004}
Yoshida, M., et al. 2004, \aj, 127, 90

\bibitem[Yoshida et al.(2008)]{Yoshida2008}
Yoshida, M., et al. 2008, \apj, 688, 918

\bibitem[Yoshida et al.(2012)]{Yoshida2012}
Yoshida, M., et al. 2012, \apj, 479, 43

\end{thebibliography}
\end{document}